\documentclass{elsart}

\usepackage{amsmath}
\usepackage{citesort}
\usepackage{epic}
\usepackage{eepic}
\usepackage{epsfig}
\usepackage{epic}
\usepackage{eepic}

\begin{document}

\runauthor{A. Kirchner and A. Schadschneider}
\begin{frontmatter}
\title{Simulation of evacuation processes using a bionics-inspired 
cellular automaton model for pedestrian dynamics}
\author[Cologne]{Ansgar Kirchner\thanksref{EmailAKI}},
\author[Cologne]{Andreas Schadschneider\thanksref{EmailAS}}
\address[Cologne]{Institut f\"ur Theoretische Physik, Universit\"at zu 
K\"oln, D-50923 K\"oln, Germany}
\thanks[EmailAKI]{E-mail: aki@thp.uni-koeln.de}
\thanks[EmailAS]{E-mail: as@thp.uni-koeln.de}

\begin{abstract}
We present simulations of evacuation processes using a recently
introduced cellular automaton model for pedestrian dynamics. 
This model applies a bionics approach to describe the interaction 
between the pedestrians using ideas from chemotaxis. 
Here we study a rather simple situation, namely the evacuation
from a large room with one or two doors.
It is shown that the variation of the model parameters allows to
describe different types of behaviour, from regular to panic.
We find a non-monotonic dependence of the evacuation times on
the coupling constants. These times depend on the strength of
the herding behaviour, with minimal evacuation times for some
intermediate values of the couplings, i.e.\ a proper combination
of herding and use of knowledge about the shortest way to the exit.
\end{abstract}

\vspace{0.7cm}
\end{frontmatter}



\section{Introduction}
Methods from physics have been successfully used for the investigation
of vehicular traffic for a long time \cite{chowd,dhrev}.
On the other hand, pedestrian dynamics has not been studied as 
extensively \cite{PedeEvak}.
Due to its generically two-dimensional nature, pedestrian motion
is more difficult to describe in terms of simple models.
However, many interesting collective effects and self-organisation 
phenomena have been observed (see \cite{dhrev,HePED} for an overview 
and a comprehensive list of references), e.g.\ jamming and clogging,
lane formation and oscillations at bottlenecks in counterflow
or collective patterns of motion at intersections. These phenomena
will be discussed in sec.~\ref{sub_CollP}.

The model takes its inspiration\footnote{Such ``learning 
from nature'' is the central idea of a field called {\em Bionics}.} 
from the process of chemotaxis (see \cite{benjacob} for a review).
Some insects create a chemical trace to guide other 
individuals to food places. This is also the central idea of the
active-walker models used for the simulation of trail formation. 
In the approach of \cite{ourpaper} the pedestrians
also create a trace. In contrast to trail formation and chemotaxis, 
however, this trace is
only virtual although one could assume that it corresponds to some
abstract representation of the path in the mind of the pedestrians.
Its main purpose is to transform effects of long-ranged
interactions (e.g.\ following people walking some distance ahead) into
a local interaction (with the ``trace''). This allows for a much more
efficient simulation on a computer. 

The basic idea of our approach might be used for studying a variety
of problems, especially from biology \cite{ourpaper,aki,debch}.
Here we want to apply this model to a simple evacuation process
with people trying to escape from a large room. Such a situation
can lead to a panic where individuals apparently act irrationally.
A nice discussion of empirical results can be found in \cite{HePED}.
Our motivation here is rather the determination and classification
of the different types of behaviour exhibited by the model than
a realistic application.

The phenomena observed during panics can be quite different from those 
found in ``normal'' situations. Nevertheless it is desirable to have a
model which is able to describe the whole spectrum of possible 
pedestrian behaviour in a unified way. So far mainly the
social-force model \cite{social} has been used which allows to
reproduce the observed behaviour \cite{dhrev,HePED,panic} quite accurately.
In this continuum model the pedestrians interact by a repulsive
(social) force which decays exponentially with the distance between
them. This means that in each step of a simulation of $N$
individuals $O(N^2)$ interaction terms have to be evaluated.
Furthermore, in complex geometries it occurs quite frequently that
two pedestrians are rather close to each other but do not interact
since they are separated by a wall. Therefore in principle one has
to check for all pairs of individuals whether an interaction is
possible or not. For large crowds this becomes very time consuming.
In contrast, in the model used here pedestrians only interact
with the floor field in their immediate neighbourhood. Therefore one
has only $O(N)$ interaction terms. A further advantage is the
discreteness of the model which allows for a very efficient 
implementation for large-scale computer simulations.

We start with a short summary of the models basic concepts.


\section{Basic principles of the model} 

\begin{figure}[h]
\begin{center}
\includegraphics[width=.5\textwidth]{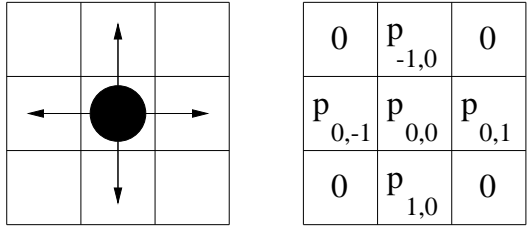}
\end{center}
\caption[]{Allowed motions and the corresponding transition probabilities.}
\label{trans}
\end{figure}

In the model the space is discretised 
into small cells which can either be empty or occupied by exactly one 
pedestrian. Each of these pedestrians can move to one of its unoccupied 
neighbour cells at each discrete time step $t\to t+1$ according to 
certain transition probabilities (see fig.~\ref{trans}). 
The probabilities are given by the interaction with two {\it floor fields} 
\cite{ourpaper}. 
These two fields $S$ and $D$ determine the transition probabilities in such 
a way that a particle movement is more likely in direction of higher fields.
They will be defined in the following subsections where also the basic
update rules of the model are summarised.


\subsection{The static floor field $S$}

\begin{figure}[h]
\begin{center}
\includegraphics[height=5cm]{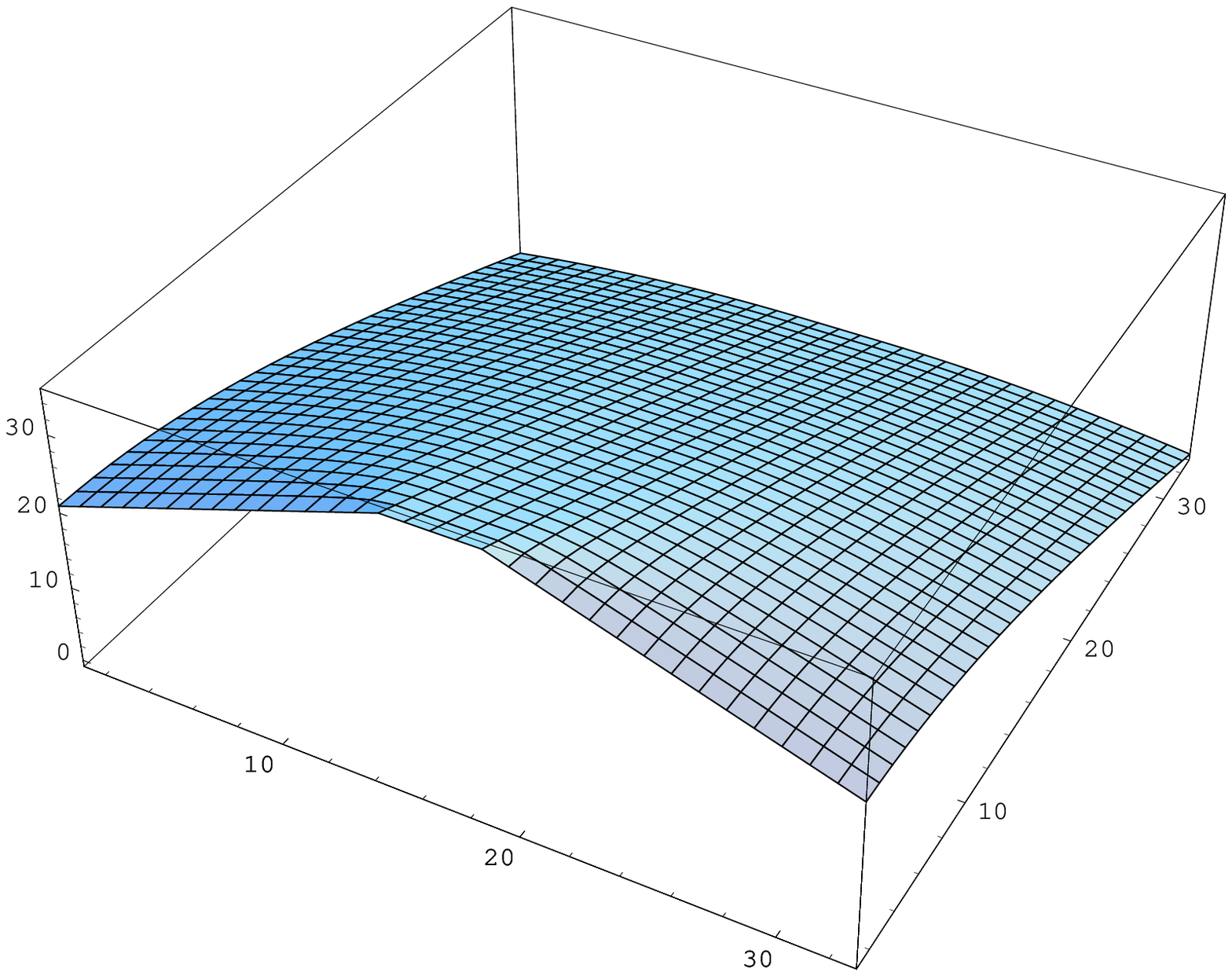}\qquad
\includegraphics[height=5cm]{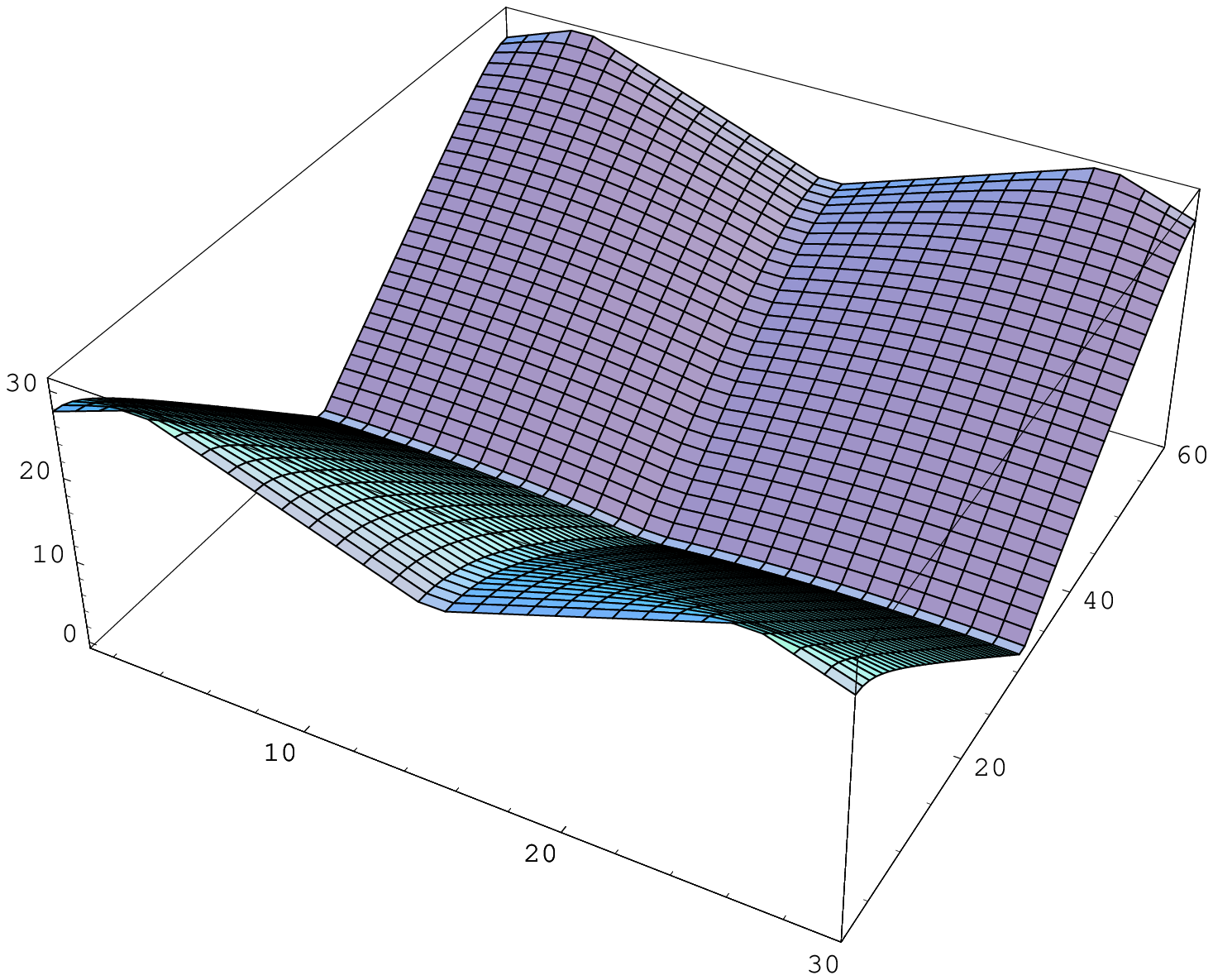}
\end{center}
\caption[]{Static floor field $S$: {\bf a)} for a lattice of size 
$X\times Y=33\times 33$ with one exit of width five 
cells; {\bf b)} lattice of size $X\times Y=30\times 60$ and four exits.}
\label{static}
\end{figure}

The static floor field $S$ does not evolve with time and is not
changed by the presence of the pedestrians. Such a field can be used
to specify regions of space which are more attractive, e.g.\ an
emergency exit or shop windows.  In case of the evacuation processes 
considered here, the static floor field describes the shortest
distance to a an exit door, lying at the middle of the top wall of the
room. Fig.~\ref{static} shows graphical representations of $S$ for
different geometries.
$S$ is calculated due to a certain distance metric for each lattice site so 
that the field values are increased in the direction to the door. The field 
values are highest for the door cells. The explicit construction of $S$ 
can be found in Appendix \ref{CoS}.


\subsection{The dynamic floor field $D$}
\label{dyn}

\begin{figure}[h]
\begin{center}
\includegraphics[height=3.5cm]{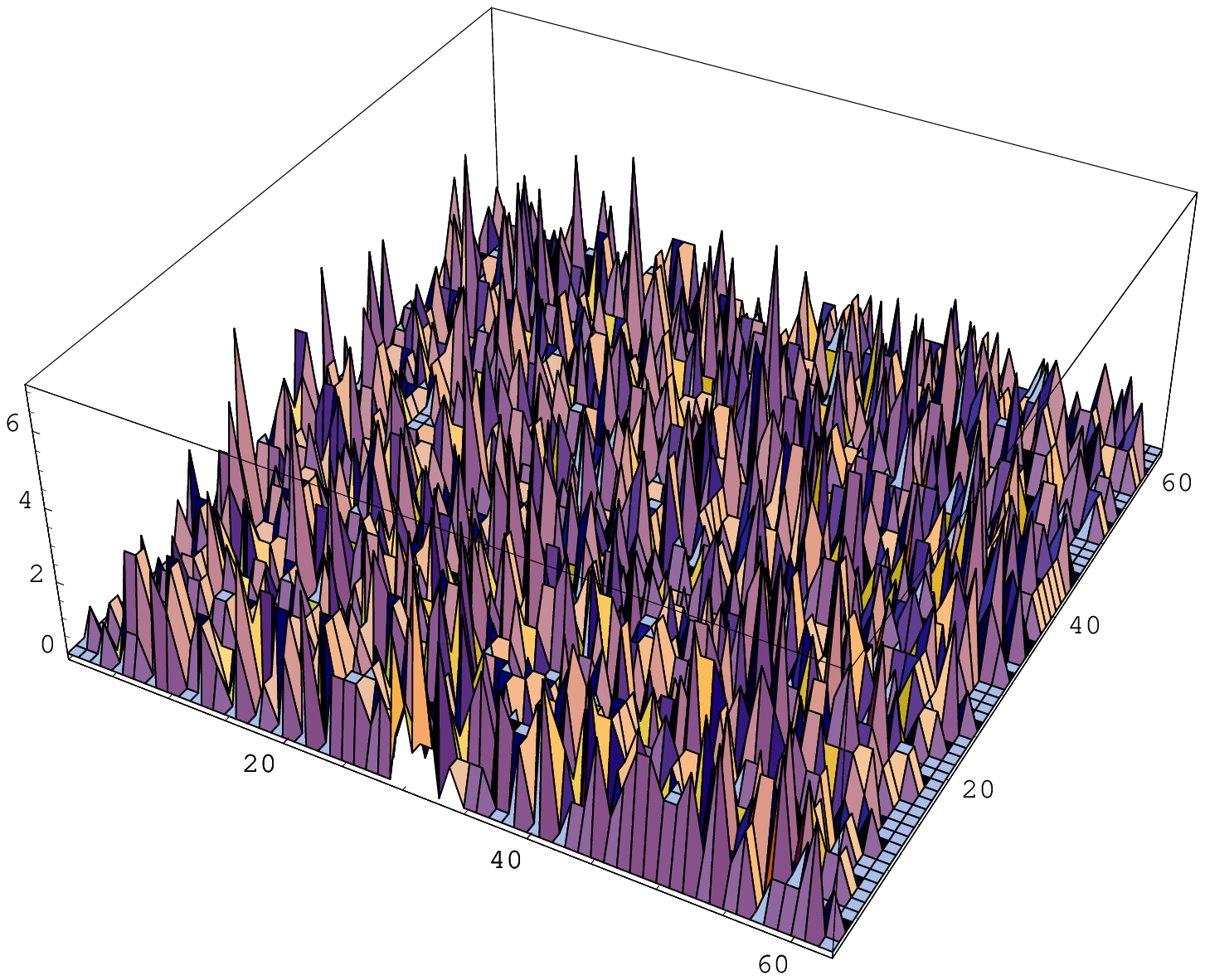}
\includegraphics[height=3.5cm]{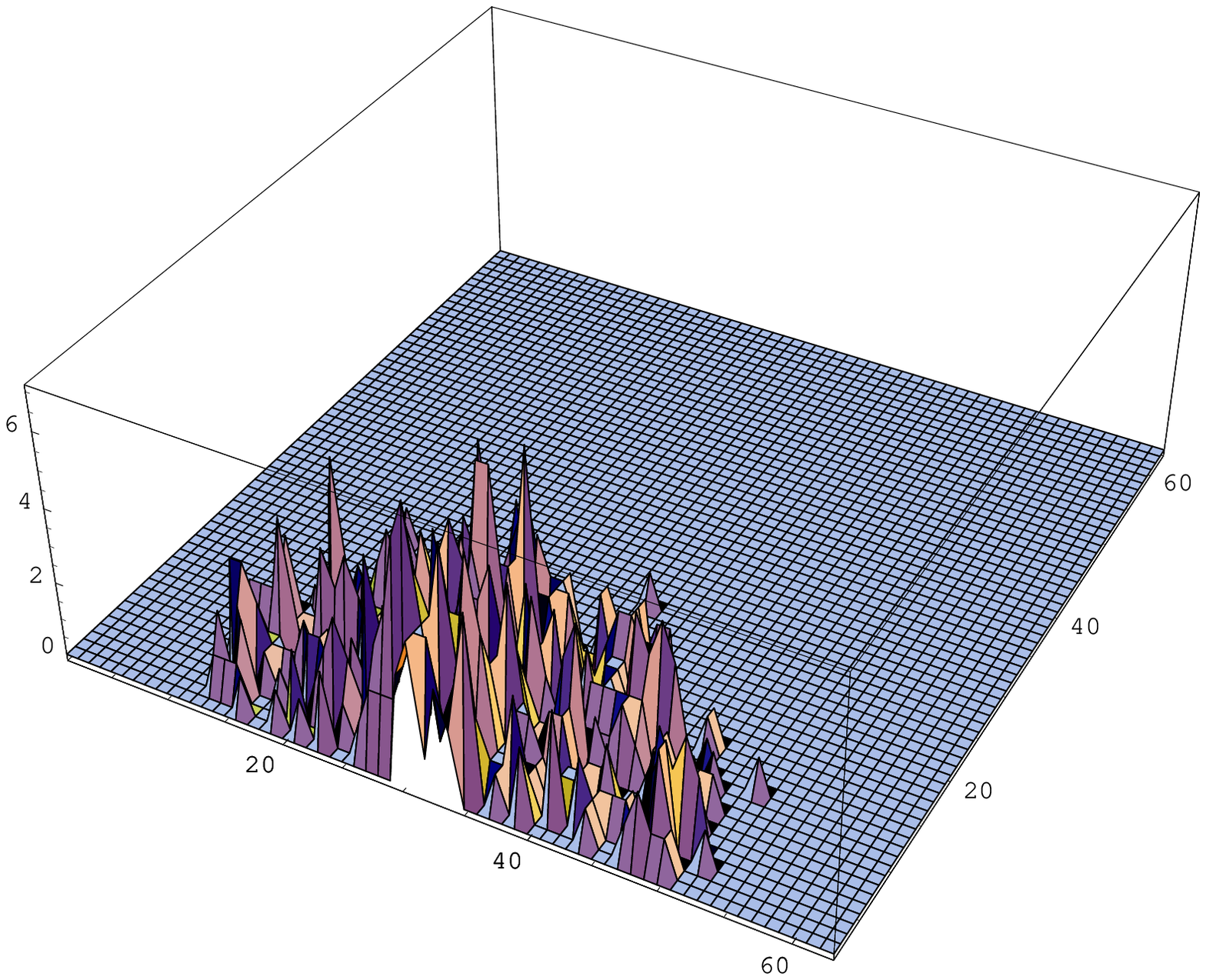}
\includegraphics[height=3.5cm]{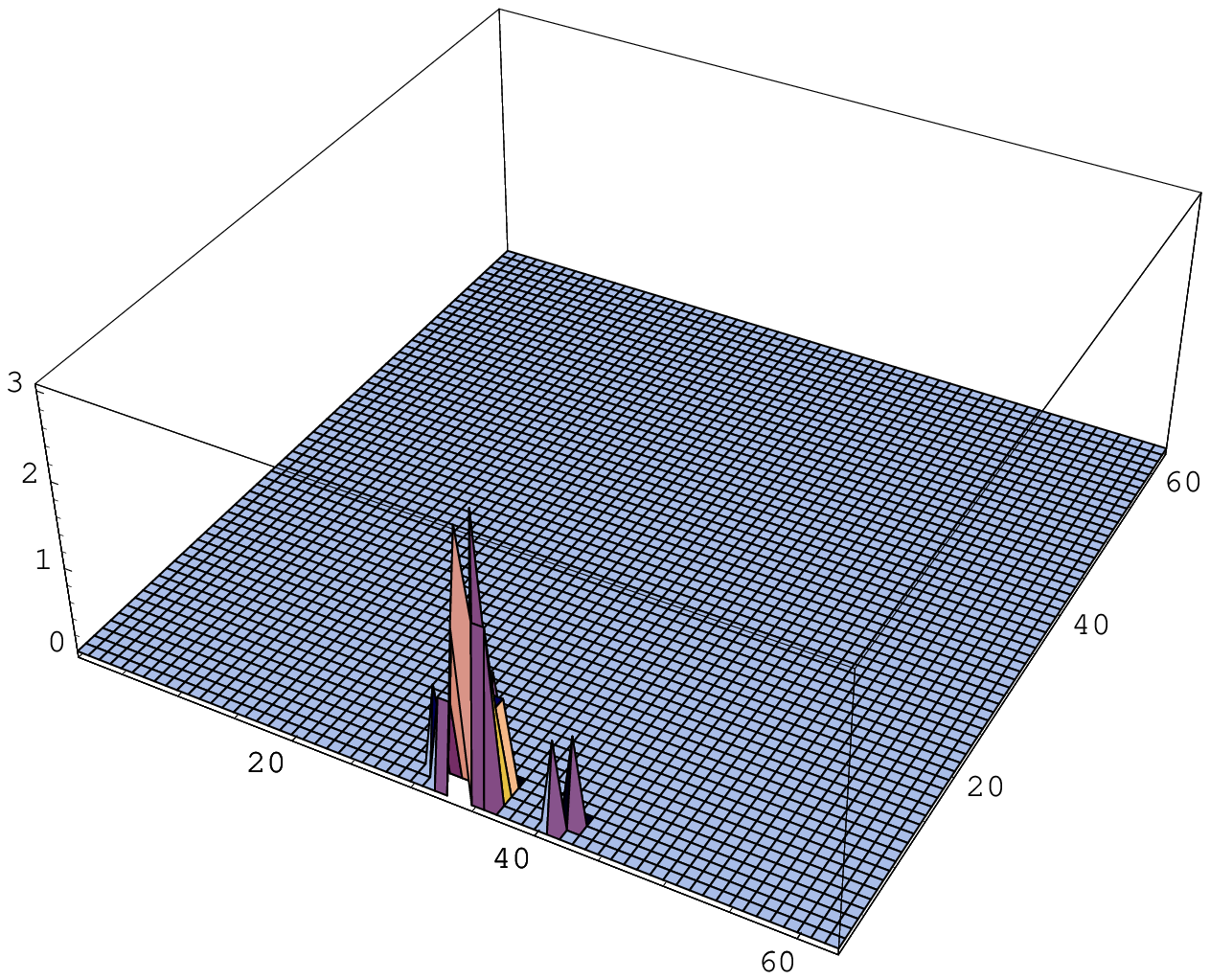}
\end{center}
\caption[]{Dynamic floor field $D$: {\bf a)} beginning evacuation process; 
{\bf b)} middle stages; {\bf (c)} final stage of evacuation.}
\label{dynamic}
\end{figure}

The dynamic floor field $D$ is a virtual trace left by the pedestrians and 
has  its own dynamics through diffusion and decay. It is used to model an 
attractive interaction between the particles. Fig.~\ref{dynamic} shows 
3-dimensional plots of $D$ for three stages during evacuation process.
At $t=0$ for all sites $(i,j)$ of the lattice the dynamic field is zero, i.e.
$D_{ij}=0$. Whenever  a particle jumps from site 
$(i,j)$ to one of  the neighbouring 
  cells,  $D$ of the starting place is increased by one: 
$D_{ij}\rightarrow D_{ij}+1$.

Therefore $D$ has only non negative integer values and can be compared to a 
{\it bosonic field}\footnote{A simplified one-dimensional model with a
fermionic field has been used recently for the description of
chemotaxis \cite{debch}.}, i.e. the bosons dropped by the pedestrians 
during their movement create the virtual trace. Thus the field value 
$D_{ij}$ corresponds to $D_{ij}$ bosons.
The dynamic floor field is time 
dependent, it has diffusion and decay controlled by two parameters
$\alpha$ $\in[0,1]$ and $\delta$ $\in[0,1]$, which means broadening and 
dilution of the trace.
In each time step of the simulation each single boson of the whole 
dynamic field $D$ decays with probability $\delta$ and diffuses with  
probability  $\alpha$ to one of its neighbouring cells.
Finally this yields $D=D(t, \delta, \alpha )$.

Note that the dynamic floor field $D$ is only altered by
moving particles and therefore it corresponds to a virtual 
{\em velocity density} field, rather then a {\em particle density} field.
In sec.~\ref{variation} we discuss an alternative definition of
the dynamic floor field $D$ as a pure particle density field and
its consequences.
In the following we give a short summary of the specific update rules used in 
the simulations.


\subsection{Update rules}

The update rules of the full model including the interaction with the 
floor fields have the following structure \cite{ourpaper,PED01a,part1}:
\begin{enumerate}
\item The dynamic floor field $D$ is modified according to its  diffusion and
 decay rules (for details, see sec.~\ref{dyn}).
\item For each pedestrian, the transition probabilities $p_{ij}$ for a move 
  to an unoccupied neighbour cell $(i,j)$ (see fig.~\ref{trans})
  are determined by the local   dynamics and the two floor fields.
  The values of the fields $D$ and $S$ are weighted with two 
  sensitivity parameters 
  $k_S\in [0,\infty[$ and $k_D\in [0,\infty[$. 
  This yields
\begin{equation}
\label{formula}
  p_{ij} = N\exp{\left(k_D D_{ij}\right)}
  \exp{\left(k_S S_{ij}\right)}(1-n_{ij})\xi_{ij}\,,
\end{equation}
with
\begin{eqnarray*}
\mbox{occupation number:}\quad && n_{ij} = 0,1,\\
\mbox{obstacle number:}\quad && \xi_{ij} = \begin{cases} 
0 & \quad \mbox{for forbidden cells, e.g.\ walls}\\
1 & \quad \mbox{else} \\
\end{cases}
,\\
\mbox{normalisation:}\quad && N = \left[\sum_{(i,j)}
\exp{\left(k_D D_{ij}\right)} \exp{\left(k_S S_{ij}\right)}(
1-n_{ij})\xi_{ij}\right]^{-1}\,. 
\end{eqnarray*}
\item Each pedestrian chooses a target cell based on the transition 
  probabilities $p_{ij}$ determined in the previous step.
\item The conflicts arising by any two or more pedestrians attempting to 
move to the same target cell are resolved by a probabilistic 
method\footnote{One can as well solve the arising conflicts using the 
procedure described in \cite{ourpaper}. It can be shown \cite{KNS}
that conflicts are important for a correct description of the
dynamics.}.
The pedestrians which are allowed to move execute their step.
\item  $D$ is increased by all moving particles.
\end{enumerate}
The above rules have to be applied to all pedestrians at the same time, 
i.e.\ we use parallel update.

Note that we do not use here the so called {\it matrix of preference} 
\cite{ourpaper} because it encodes a direction of preferred motion of the 
pedestrians, which they usually not have at the beginning of an evacuation 
process. In the following all information about the desired direction of 
motion is obtained from the floor fields.

In \cite{ourpaper} it has been shown that simple, local update rules 
for the pedestrians and the two floor fields are sufficient to yield 
a richness of complex phenomena (see sec.~\ref{sub_CollP}). 
Obviously this route is superior concerning the computational efficiency and 
even allows for faster-than-real-time simulations of large crowds
\cite{ourpaper,PED01b,part2}, e.g.\ in evacuation processes in 
public buildings.


\subsection{Collective Phenomena}
\label{sub_CollP}

Pedestrian dynamics exhibits are variety of fascinating collective
effects \cite{dhrev,HePED}:

\begin{itemize}
\item {\bf Jamming}: 
At large densities various kinds of jamming phenomena occur, e.g.\
when many people try to leave a large room at the same time
\cite{HePED,ourpaper,PED01b,panic,part2,tajima}.
This clogging effect is typical for a bottleneck situation.
Other types of jamming occur in the case of counterflow where two groups 
of pedestrians mutually block each other \cite{fukui,nagatani99}. 
This happens typically at high densities and when it is not possible 
to turn around and move back, e.g.\ when the flow of people is large.

\item {\bf Lane formation}: In counterflow, i.e.\ two groups of
people moving in opposite directions, a kind of spontaneous symmetry
breaking occurs. 
The motion of the pedestrians can self-organise
in such a way that (dynamically varying) lanes are formed where
people move in just one direction \cite{social}. In this way, strong 
interactions with oncoming pedestrians are reduced and a higher walking 
speed is possible.

\item {\bf Oscillations}: In counterflow at bottlenecks, e.g.\ doors, one 
can observe oscillatory changes of the direction of motion \cite{social}.
Once a pedestrian is able to pass the bottleneck it becomes easier
for others to follow him in the same direction until somebody is
able to pass (e.g.\ through a fluctuation) the bottleneck in the
opposite direction.

\item{\bf Patterns at intersections}: At intersections various collective
patterns of motion can be formed \cite{dhrev,HePED,nagatani}. A typical 
example are short-lived roundabouts \cite{dhrev,HePED}
which make the motion more efficient. 
Even if these are connected with small detours the formation of these 
patterns can be favourable since they allow for a ``smoother'' motion.

\item {\bf Trail formation}: Although human and animal trails
are formed for rather different purposes their structures
have some similarities \cite{activewalker,trail}. Often
human trails are formed as a short-cut which makes it attractive
to leave a paved path. Animal trails usually are related to
chemotaxis and mark the way to food places.

\item{\bf Panics}: In panic situations many counter-intuitive 
phenomena (e.g.\ ``faster-is-slower'' and ``freezing-by-heating'' 
effects \cite{HFV}) can occur. For a thorough discussion we refer
to \cite{HePED,panic} and references therein. 
\end{itemize}

In \cite{ourpaper,PED01b} it has been shown that the new CA model 
described in the previous subsection is  -- despite its simplicity --
able to reproduce these observed collective effects. 
This is essential if one intends to use
the model for real applications, e.g.\ the optimisation of evacuation
procedures.


\section{Evacuation Simulations}

In the following we describe results of simulations of a typical situation, 
i.e. the evacuation of a large room (e.g.\ in case of fire). At this  we focus 
on  the influence of the sensitivity parameters $k_D$ and $k_S$ on the 
evacuation times in order to identify the different classes of
behaviour exhibited by the model. As we will see 
interesting collective phenomena between the pedestrians lead to a
nontrivial dependence of the evacuation times on $k_D$ and $k_S$.
In sec.~\ref{variation} we investigate possible alternative definitions
of the dynamical floor field. Sec.~\ref{twodoors} is devoted to a
simple optimisation problem, namely evacuation from a room with two 
doors \cite{webpage}.


\subsection{The impact of sensitivity parameters}
\label{sub_sens}

The value of $k_S$, the coupling to the static field, can be viewed as
a measure of the knowledge of the pedestrians about the location of
the exit. A large $k_S$ implies a motion to the exit on the shortest
possible path. For vanishing $k_S$, on the other hand, the people will
perform a random walk and just find the door by chance. So the case 
$k_S\ll 1$ is relevant for processes in dark or smoke-filled rooms
where people do not have full knowledge about the location of the exit.

The parameter $k_D$ for the coupling to the dynamic field controls the
tendency to follow the lead of others. A large value of $k_D$ implies
a strong herding behaviour which has been observed in the case of
panics \cite{panic}.

\begin{figure}[ht]
\begin{center}
\includegraphics[width=0.3\textwidth]{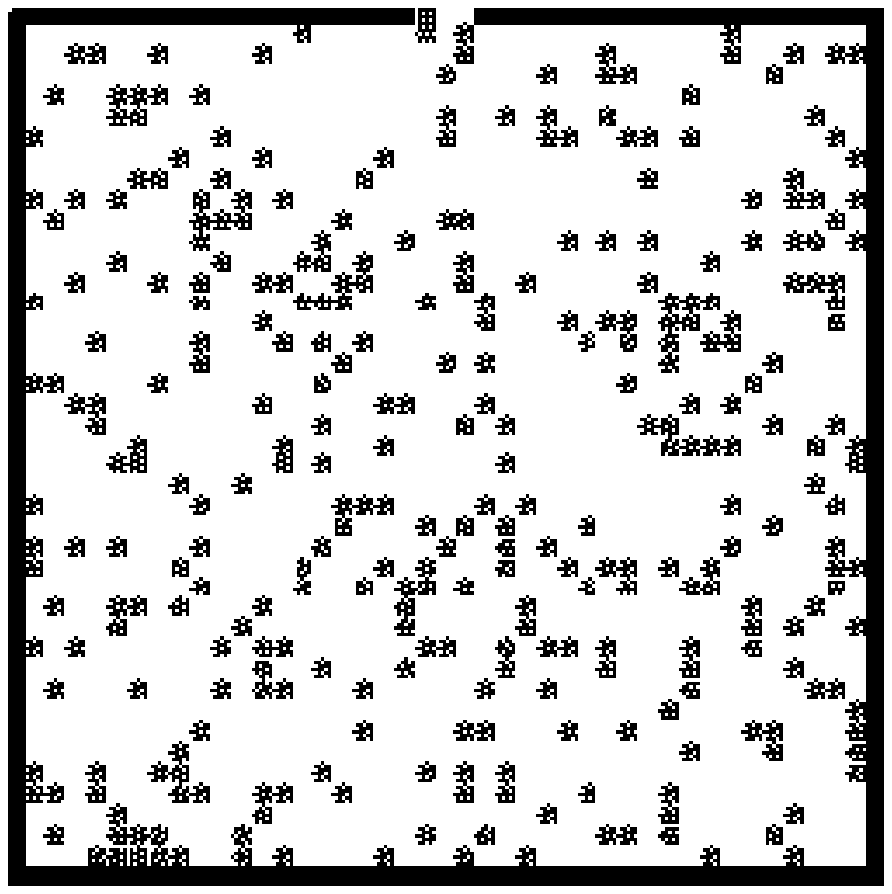}
\includegraphics[width=0.3\textwidth]{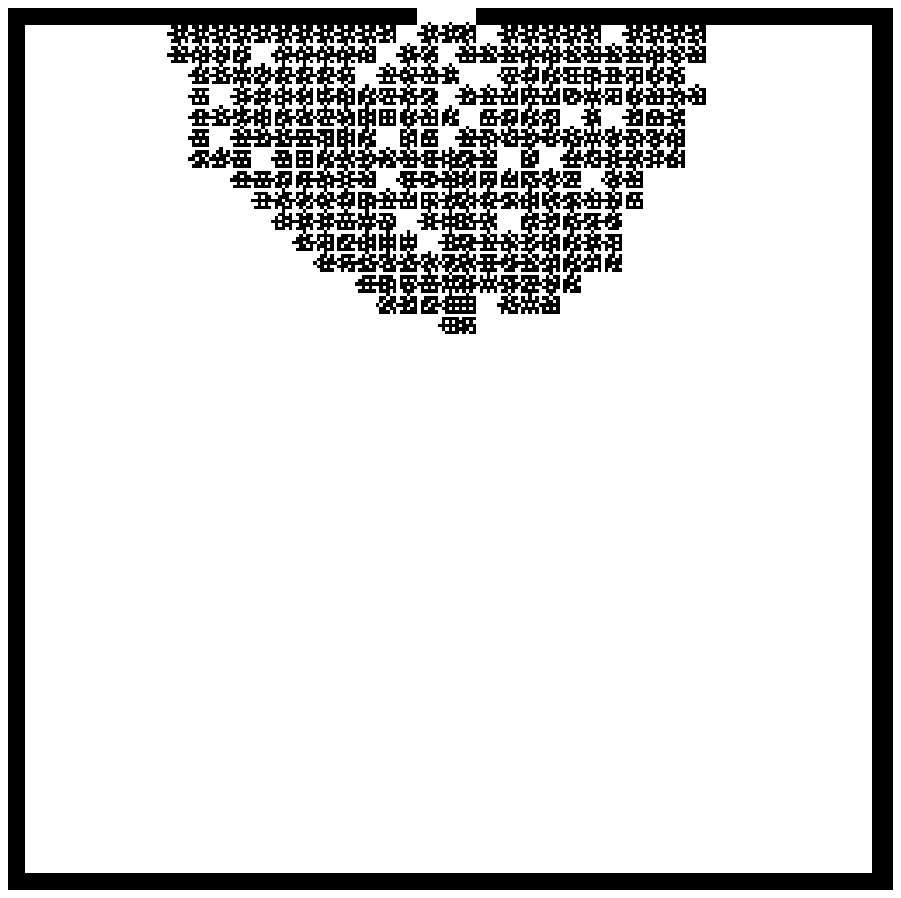}
\includegraphics[width=0.3\textwidth]{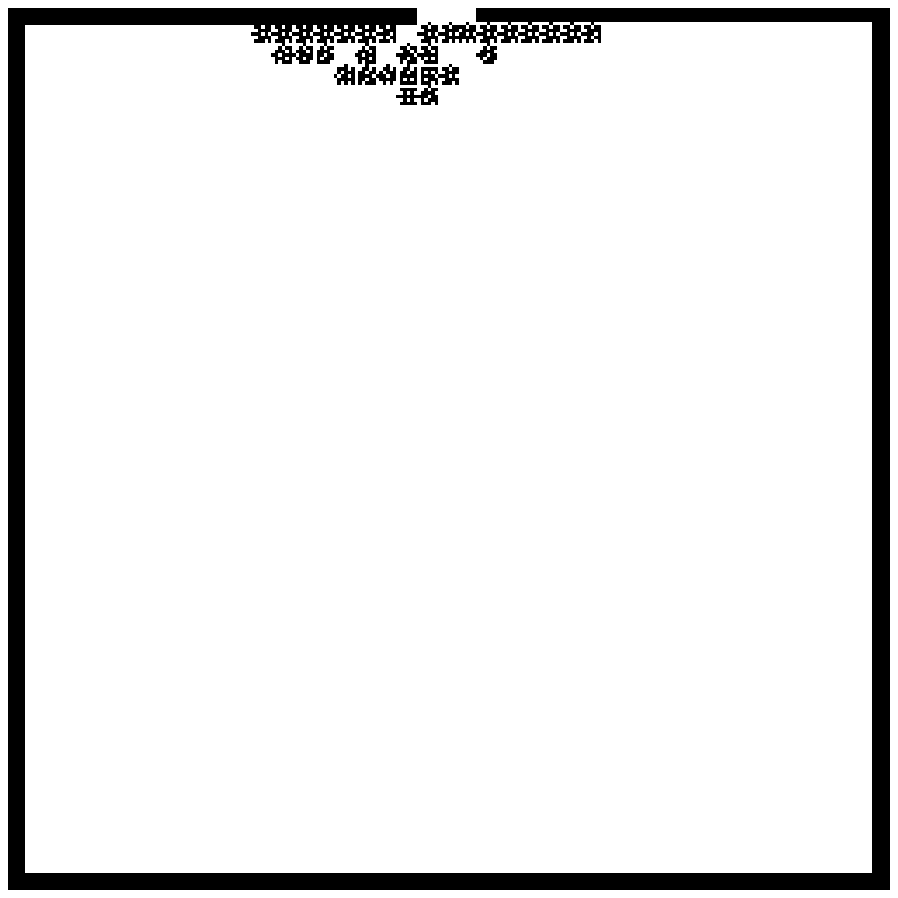}
\end{center}
\caption[]{Typical stages of the dynamics: {\bf (a)} beginning evacuation 
($t=0$); {\bf (b)} middle stages; {\bf (c)} end stage of evacuation with only 
a few particles left.}
\label{snap}
\end{figure}

We consider a grid of size $63\times 63$ sites with an exit of one cell in 
the middle of one wall. The particles are initially distributed randomly
 and try to leave the room. The only information they get is through the 
floor fields. Fig.~\ref{snap} shows typical stages of the dynamics for an 
initial particle density of $\rho=0.3$, which means 1116 particles.
In the middle picture of fig.~\ref{snap} a half-circle
jamming configuration in front of the door is easy to spot. 
A typical feature of the dynamics is a radial motion of `holes'
created by particles escaping through the door.

\begin{figure}[h]
\begin{center}
\includegraphics[width=0.49\textwidth]{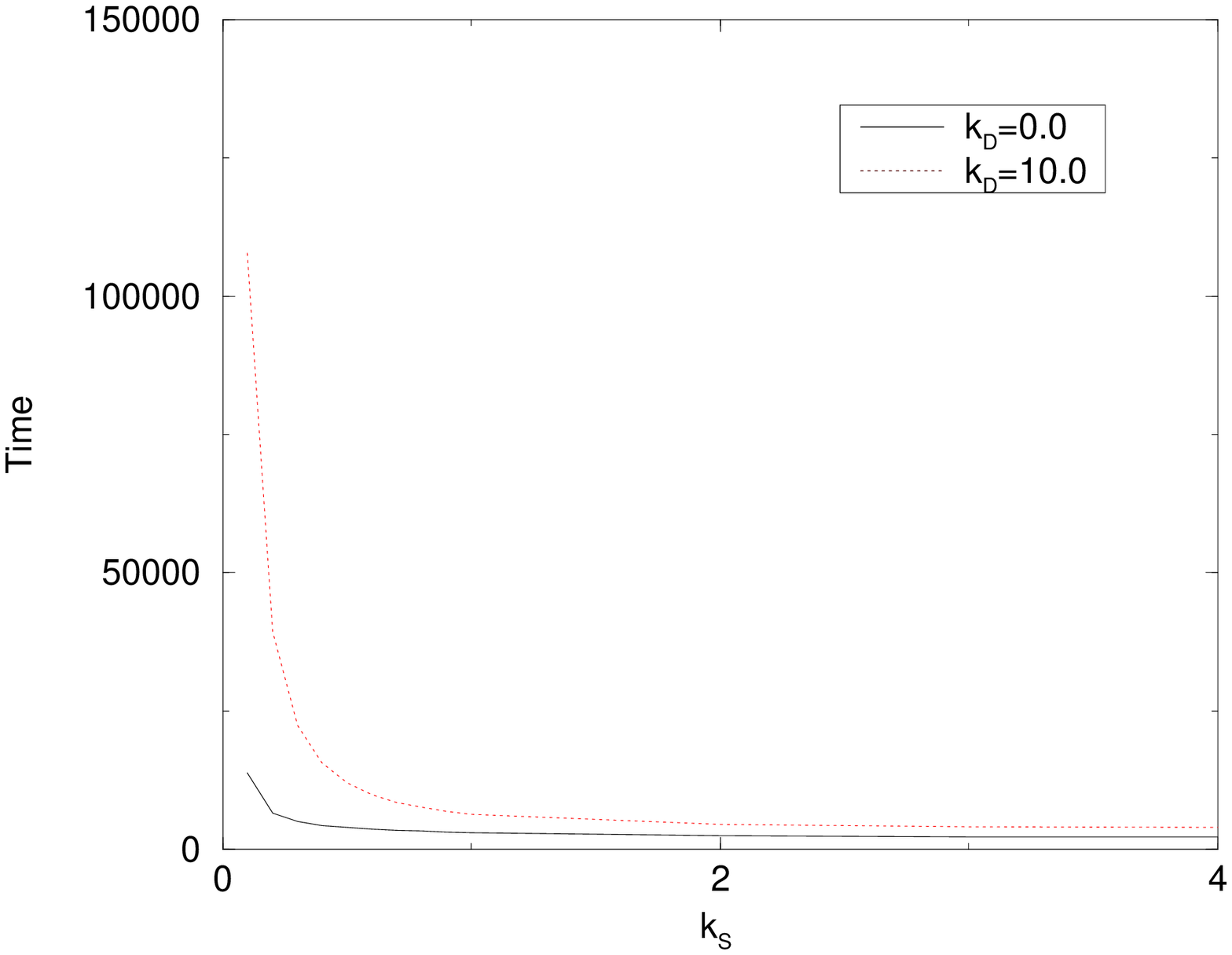}
\includegraphics[width=0.49\textwidth]{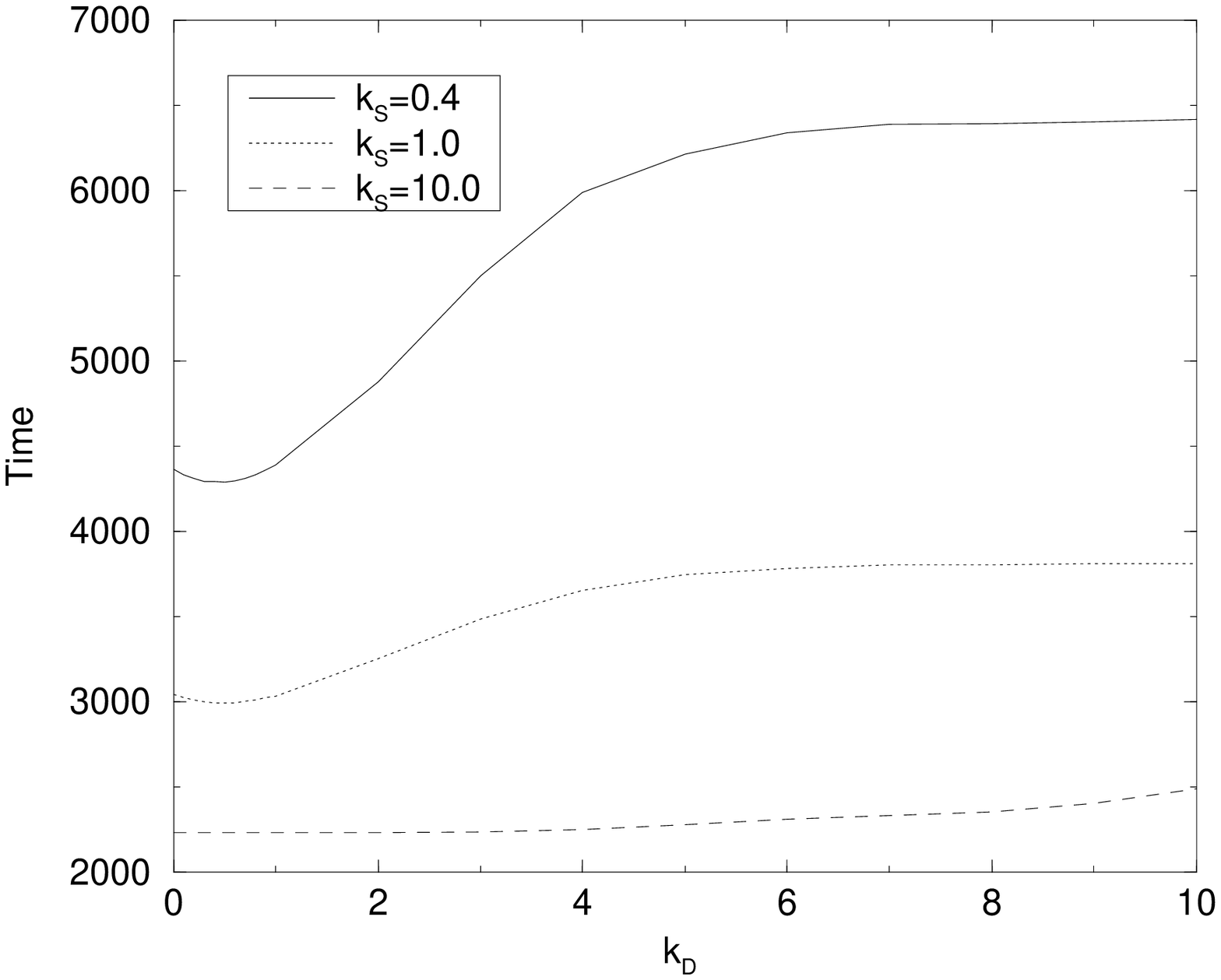}
\end{center}
\caption{Averaged evacuation times for a large room (fig.~\ref{snap}) with
 an initial particle density of $\rho=0.3$ and decay constant $\delta =0.3$: 
{\bf (a)} $\alpha=0.1$ and fixed $k_D$;
{\bf (b)} $\alpha=0.3$ and fixed $k_S$.}
\label{fixed_kdks}
\end{figure}

\begin{figure}[h]
\begin{center}
\includegraphics[width=0.49\textwidth]{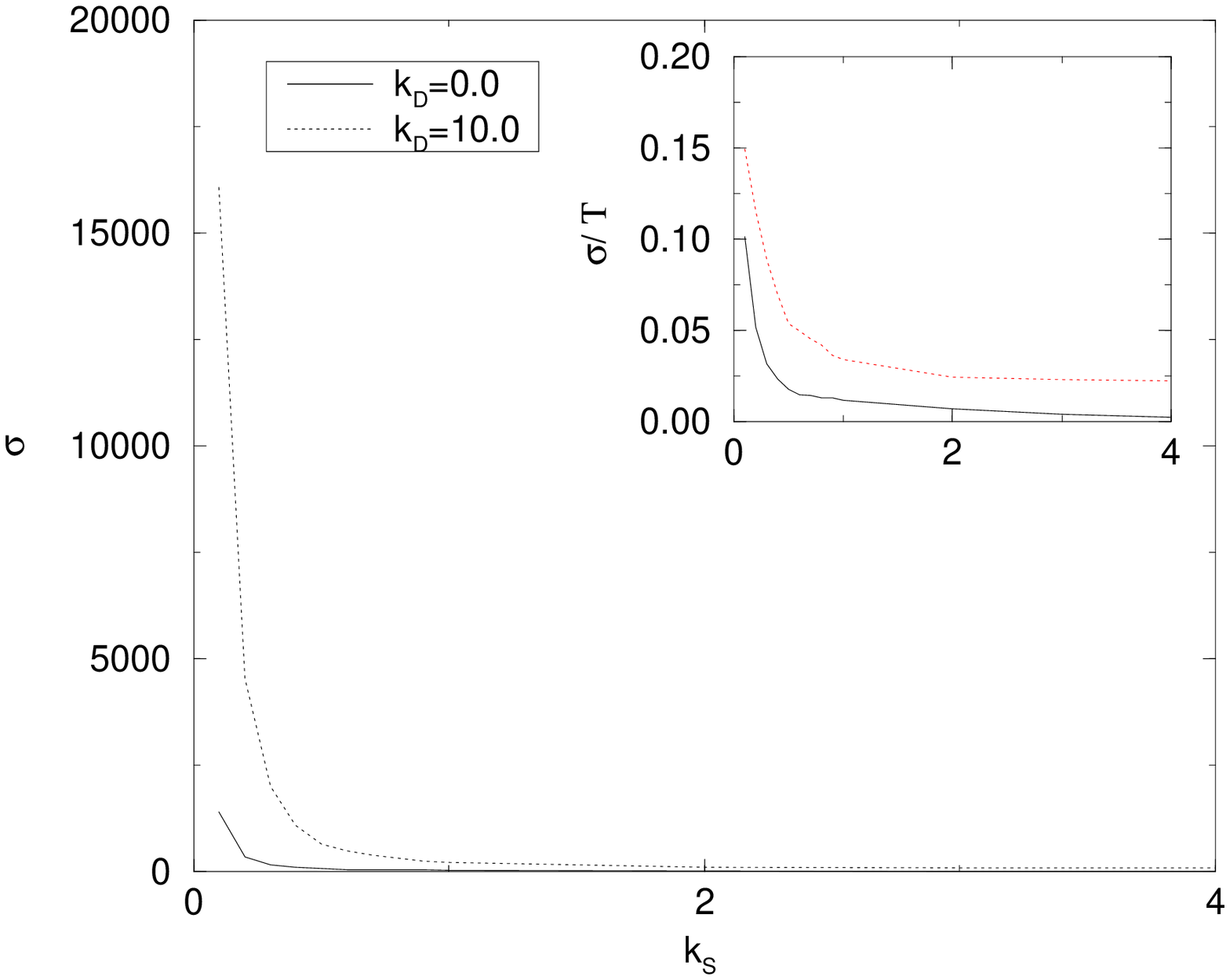}
\includegraphics[width=0.49\textwidth]{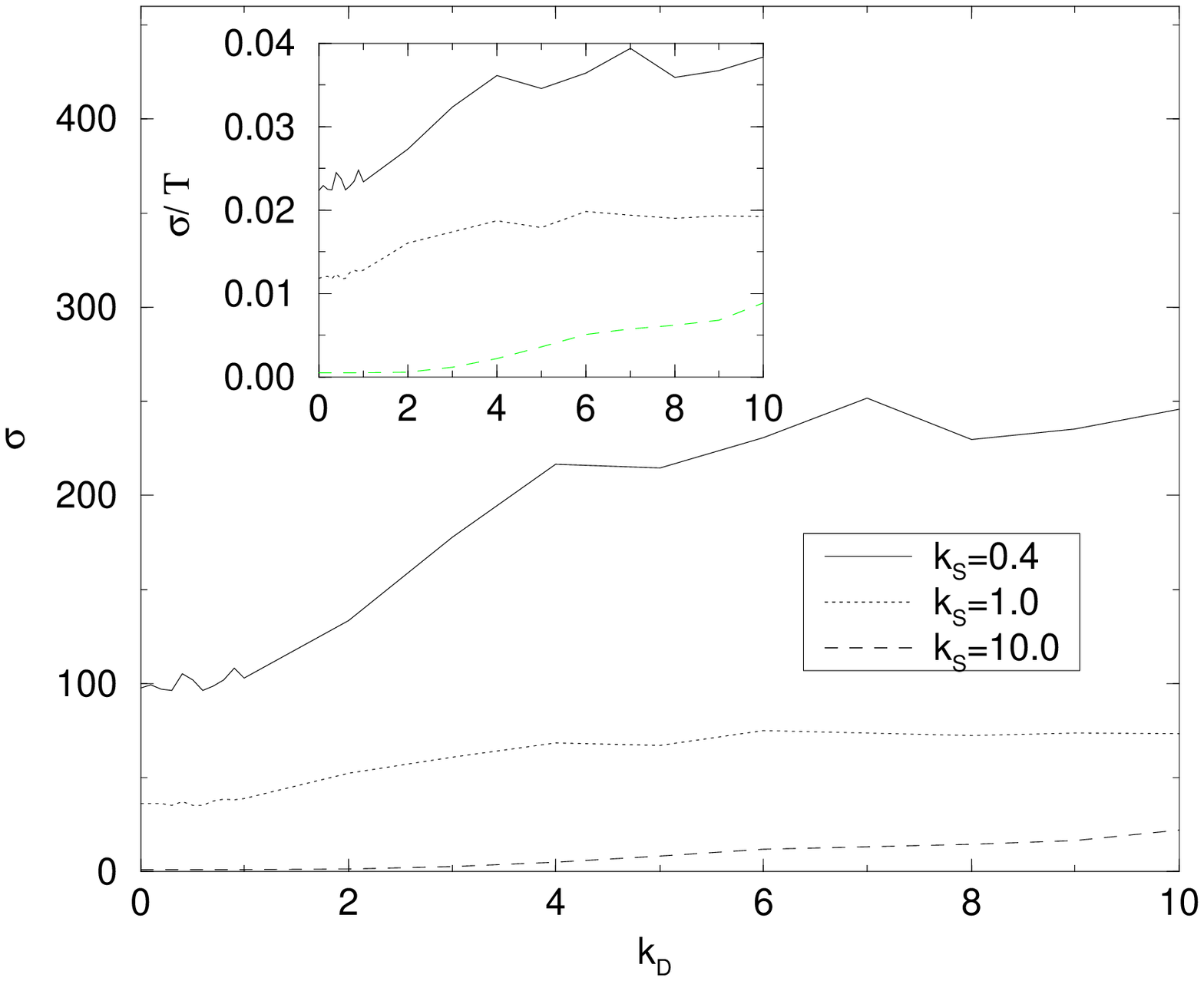}
\end{center}
\caption{Averaged variances $\sigma$ and relative variances $\sigma/T$ of 
the  evacuation times of Fig.~\ref{fixed_kdks}: {\bf (a)} $\delta =0.3$, 
$\alpha=0.1$ and fixed $k_D$;
 {\bf (b)} $\delta =0.3$, $\alpha=0.3$ and fixed $k_S$.}
\label{fixed_kdks_var}
\end{figure}

The evacuation times $T$ and their variances $\sigma$ are averaged over 
$500$ samples and strongly
depend on the sensitivity parameters $k_D$ and $k_S$.
Fig.~\ref{fixed_kdks} (a) shows the evacuation times for fixed sensitivity
parameter $k_D$ of the dynamic field and variable sensitivity
parameter $k_S$ of the static field and  fig.~\ref{fixed_kdks} (b) shows the evacuation times for fixed $k_D$  and variable $k_S$. Fig.~\ref{fixed_kdks_var} shows all 
corresponding variances $\sigma$ and relative variances $\sigma/T$ of 
 the evacuation times of fig.~\ref{fixed_kdks}.
The averaged evacuation times are always measured in update time steps. 
With the generic time-scale of the model \cite{ourpaper,PED01a,part1} of
about $0.3$ sec/timestep it easy to translate that into a real time value.

Let us first consider the case of $k_D=0$, i.e.\ no coupling to the 
dynamic field.
In fig.~\ref{fixed_kdks}(a) one can see the influence of $k_S$. 
For $k_S\rightarrow 0$ the pedestrians do not sense the strength of the 
field. Therefore they do not have any guidance through the surroundings 
and perform a pure random walk which leads to a maximal value 
of the evacuation time for $k_S=0$. 
For $k_S\rightarrow \infty$ they have full information about the shortest
distance to the door and the evacuation time converges towards a minimal
value. The movement of the particles becomes almost deterministic. 
Therefore $k_S$ can be interpreted as some kind of inverse temperature
for the degree of information about the inanimate surrounding of the
pedestrians.

In the same way the sensitivity parameter $k_D$ of the dynamic field
works as an inverse temperature for the information about the virtual
trace. If $k_S$ is turned on from zero to infinity, a non-zero value
of $k_D$ only means additional noise to the pedestrians and
evacuation times increase for higher coupling strength to $k_D$
(see fig.~\ref{fixed_kdks}(a)).

Much more interesting is the behaviour for fixed $k_S$
(fig.~\ref{fixed_kdks}(b)). 
The evacuation times saturate at maximal values for 
growing sensitivity parameter $k_D$ of the dynamic field. 
The most interesting point is non-monotonic behaviour of $T(k_D)$ with 
the occurrence of minimal evacuation times for non-vanishing small 
values of the sensitivity parameter $k_D$ of the dynamic field. 
Therefore a small interaction with the dynamic field, which is proportional 
to the velocity density of the particles,
represents some sort of minimal intelligence of the pedestrians.  
They are able to detect regions of higher local flow and minimise 
their waiting times. 
This effect is most pronounced for intermediate coupling  ($k_S\approx 0.4$)
to the static field $S$, but very weak for strong coupling to $S$ (e.g.\ 
$k_S\approx 10$), where the evacuation is dominated by the orientation to 
the inanimate surrounding. It vanishes again for very weak coupling to 
$S$ ($k_S=0.1$, not shown in fig.~\ref{fixed_kdks}(b)) where the movement 
of the pedestrians is similar to a random walk.

If the coupling to the dynamic field is further increased, the
evacuation times increase again 
and saturate at maximal values. The interaction with other pedestrians 
becomes more and more unfavourable similar to the arising of a panic 
situation.  The weaker the coupling to the static field $S$ is,
the higher are the evacuation times: the particles than have less
information of the inanimate surrounding (for example they cannot find
the way to the door because of smoke in a fire situation).

The variances $\sigma$ of the evacuation times (fig.~\ref{fixed_kdks_var}) 
show qualitatively the same behaviour as
the averaged time values. For strong coupling to $S$ 
($k_S\approx 10$, $k_D<2$) $\sigma$ tends to zero, which means that the 
process is nearly deterministic. 
For weak coupling to $S$ the variance $\sigma$ increases strongly and
also the relative variance $\frac{\sigma}{T}$ becomes rather large
for $k_S\to 0$. 
From a practical point of view, i.e.\ evacuations of real buildings,
one has to ask whether it makes sense to specify safety just by the
average evacuation time alone when (relative) variances can become
large.
For fixed $k_S$ and varying coupling $k_D$ to the dynamic field the
variance $\sigma$ also behaves qualitatively as $T$, i.e.\ it increases
with increasing $k_D$ (see fig.~\ref{fixed_kdks_var}(b)).

In the following we discuss the influence of the decay and diffusion 
parameters $\delta$ and $\alpha$.
\begin{figure}[ht]
\begin{center}
\includegraphics[width=0.49\textwidth]{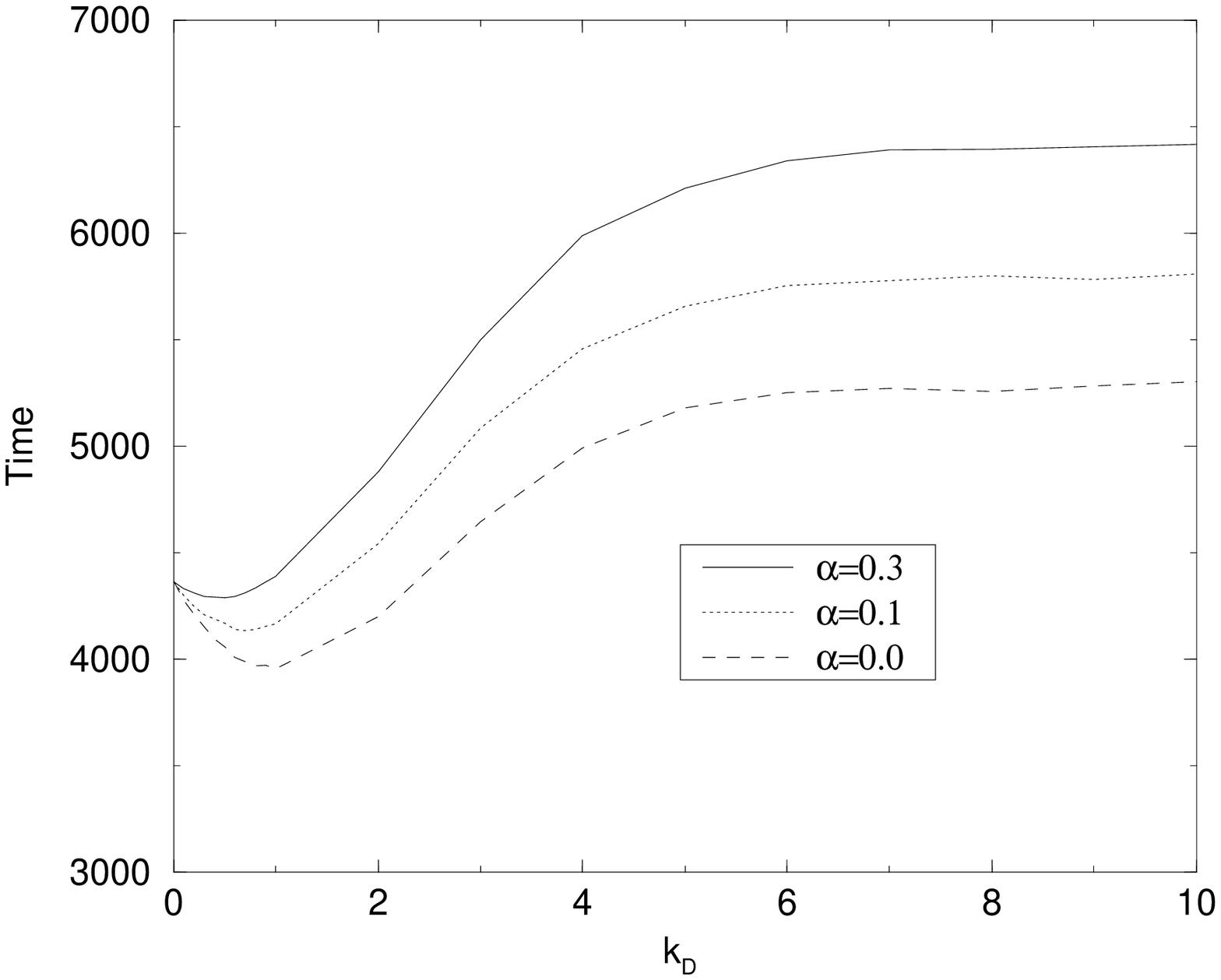}
\includegraphics[width=0.49\textwidth]{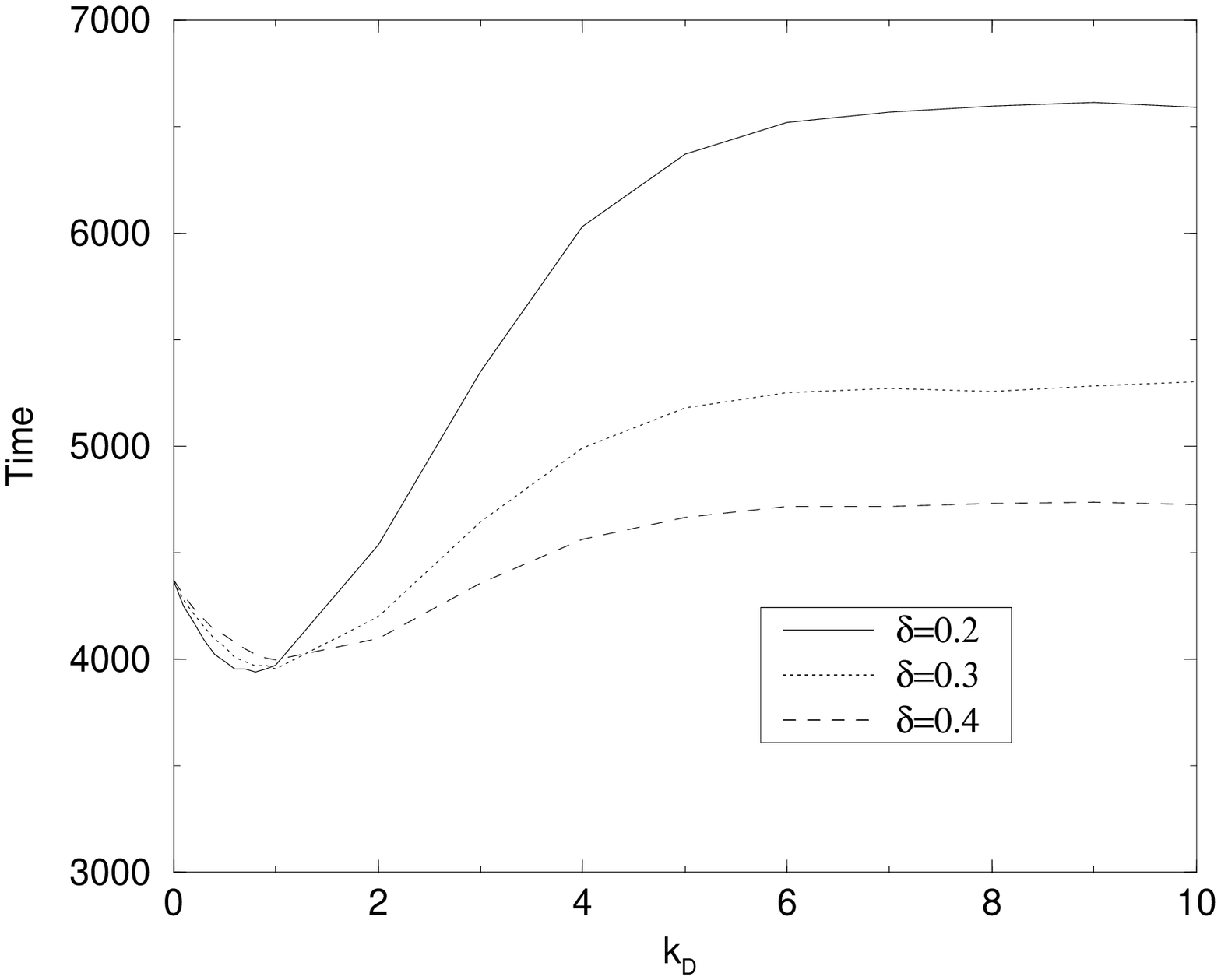}
\end{center}
\caption{Averaged evacuation times for a large room for an
 initial particle density of $\rho=0.3$: {\bf (a)} $k_S=0.4$ and 
$\delta=0.3$; {\bf (b)} $k_S=0.4$ and $\alpha=0.0$.}
\label{alphadelta}
\end{figure}
In fig.~\ref{alphadelta} (a) one can see that for $k_S=0.4$ and a high 
density of $\rho=0.3$ the effect of minimal evacuation times for 
non-vanishing small values of $k_D$ 
becomes most pronounced in the limit $\alpha\to 0$. Besides that all 
evacuation times are increased with increasing $\alpha$. That implies
that diffusion of the field only means additional noise and no advantageous 
information for the particles due to the fast broadening and dilution 
of the trace for large $\alpha$.

For decay parameter $\delta\rightarrow 0$ the dynamic field persists for 
a long time and the regime of minimal evacuation times is shifted towards 
smaller $k_D$ values.
Evacuation times for fixed $k_D$ increase monotonically with $\delta$
(see fig.~\ref{alphadelta}(b)).
This is a collective effect where the herding behaviour helps to overcome
the insufficient knowledge about the location of the exit.
In contrast, the evacuation times in the panic regime (i.e.\ high $k_D$ 
values) are increased with smaller $\delta$. With the field values of $D$ 
the memory of previous steps grows which increases the noise 
in the system for the particles.  Therefore for large values of $k_D$ 
a strong herding behaviour would be unfavourable since it tends to
`confuse' the pedestrians which already have a good knowledge about the
geometry. This reversed behaviour in comparison with the case $k_D\to 0$
leads to a crossing of the curves in fig.~\ref{alphadelta}(b).
It is interesting to note that this crossing happens in a rather narrow
interval around $k_D\approx 1$.

The influence of the diffusion parameter $\alpha$ of the dynamic field
in the regime of minimal evacuation times strongly depends on the
particle density $\rho$. 
\begin{figure}[ht]
\begin{center}
\includegraphics[width=0.49\textwidth]{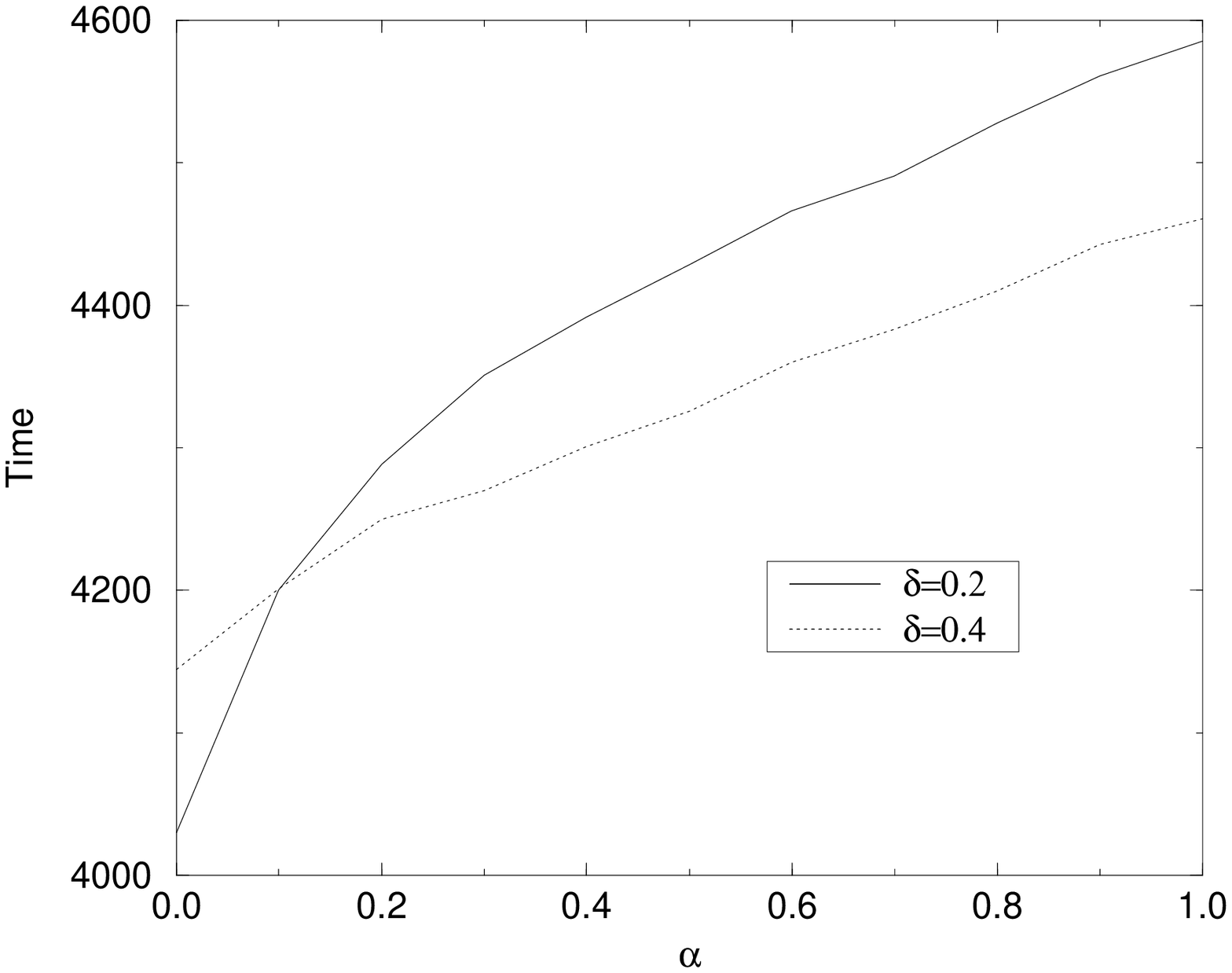}
\includegraphics[width=0.49\textwidth]{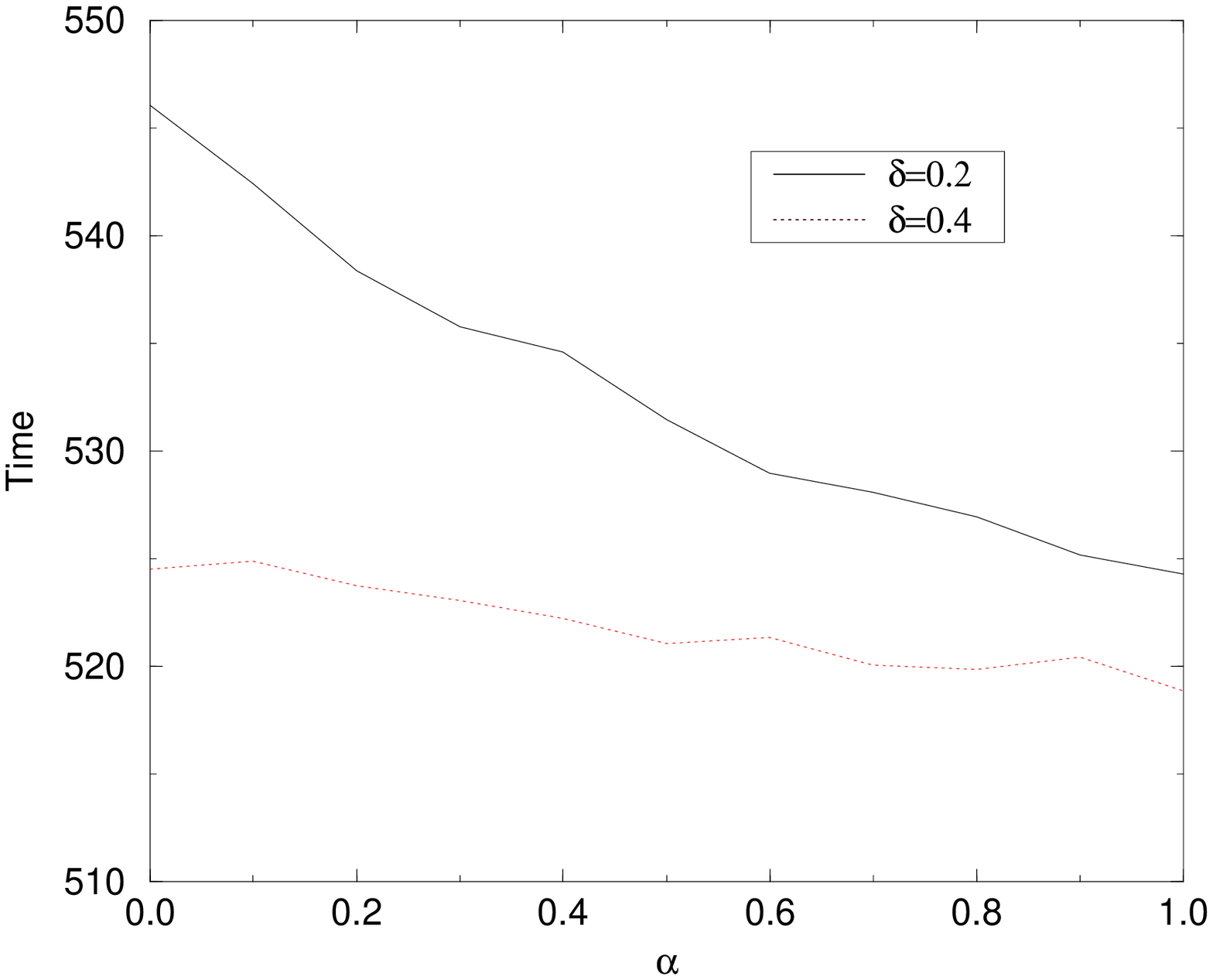}
\end{center}
\caption{Averaged evacuation times for a large room (Fig.~\ref{snap}) for 
two densities, variable $\alpha$ and $k_S=k_D=0.4$: {\bf (a)} $\rho=0.3$,
and {\bf (b)} $\rho=0.003$.}
\label{diff}
\end{figure}
As in fig.~\ref{alphadelta}(a) in  fig.~\ref{diff}(a) one can see that 
in the high density regime 
an increasing diffusion parameter $\alpha$ always increases 
the evacuation times for arbitrary decay parameter $\delta$, since
the diffusion of the field increases the noise for the particles.
In the low density regime one finds the opposite behaviour.
Here the diffusion of the dynamic field leads to a 
higher degree of information for the particles and to favourably long-ranged 
interactions between them (fig.~\ref{diff}(b)).
For large $\alpha$ the evacuation time increases with decreasing $\delta$
whereas for small $\alpha$ one finds just the opposite behaviour.
This leads to a crossing of the curves in fig.~\ref{diff}(a).
The main reason is that for larger densities the evacuation time is
mostly determined by the clogging which occurs at the exit whereas
for small densities clogging is negligible. Furthermore for
small $\alpha$ the traces are rather sharp and
so a larger $\delta$ just reduces the field strength. In contrast, for
large $\alpha$ the traces broaden quickly and are thus diluted such
that they form an almost constant background.

Thus three main regimes for the behaviour of the particles can be 
distinguished. For strong coupling to $k_S$ and very small coupling 
to $k_D$ we find an {\it ordered regime} where particles only react to 
the static floor field and the behaviour than is in some sense 
deterministic.
The {\it disordered regime}, characterised by strong coupling to $k_D$ 
and weak coupling to $k_S$, leads to a maximal value of the evacuation 
time. The behaviour here is typical for panic situations, e.g.\ 
the herding tendency dominates.
Between these two regimes an {\it optimal regime} exists where the
combination of interaction with the static and the dynamic floor
fields minimises the evacuation time. Here the individuals have
some knowledge about the location of the exits, but through some sort 
of cooperation the evacuation time $T$ can be optimised.

\begin{figure}[ht]
\begin{center}
\includegraphics[width=0.49\textwidth]{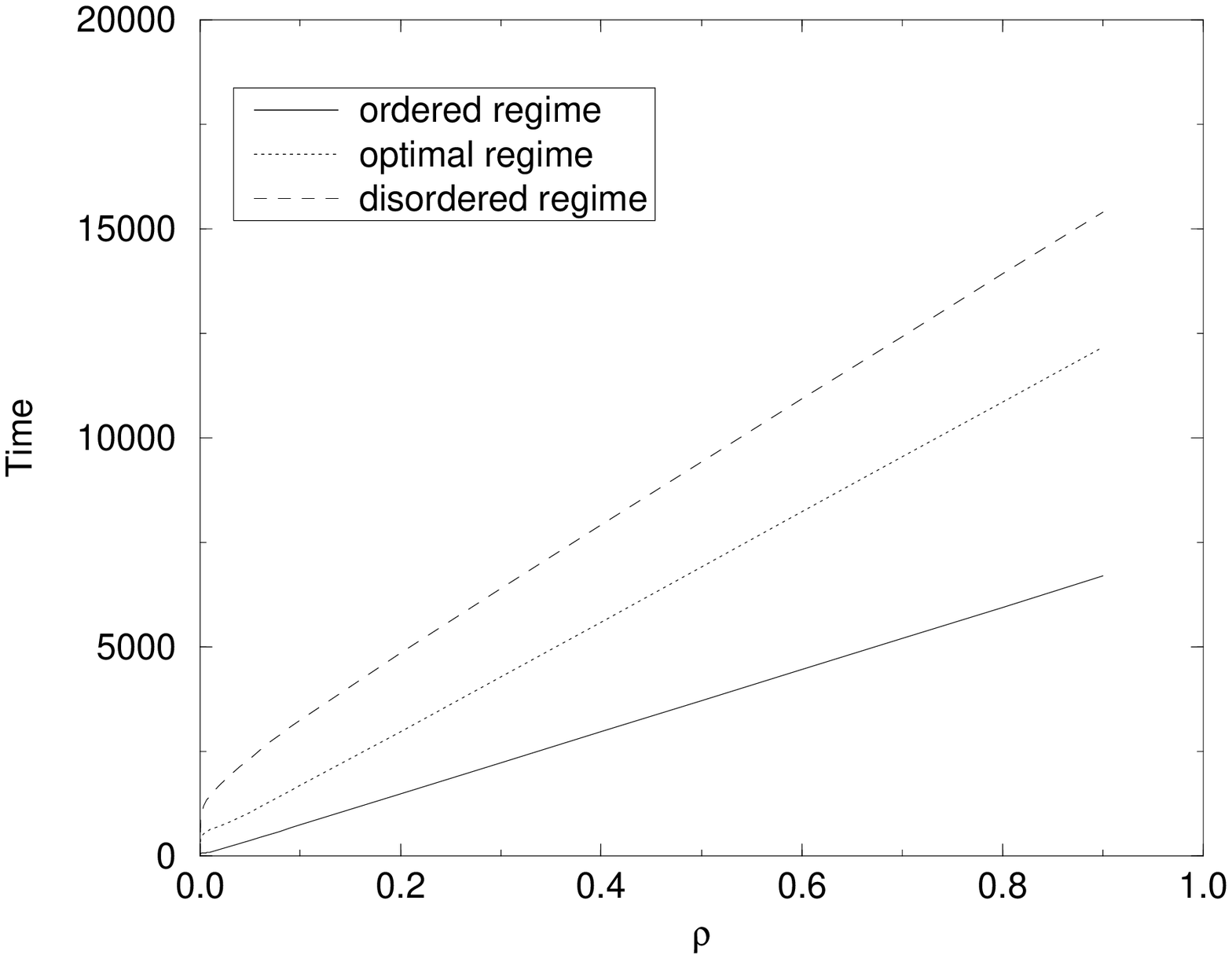}
\includegraphics[width=0.49\textwidth]{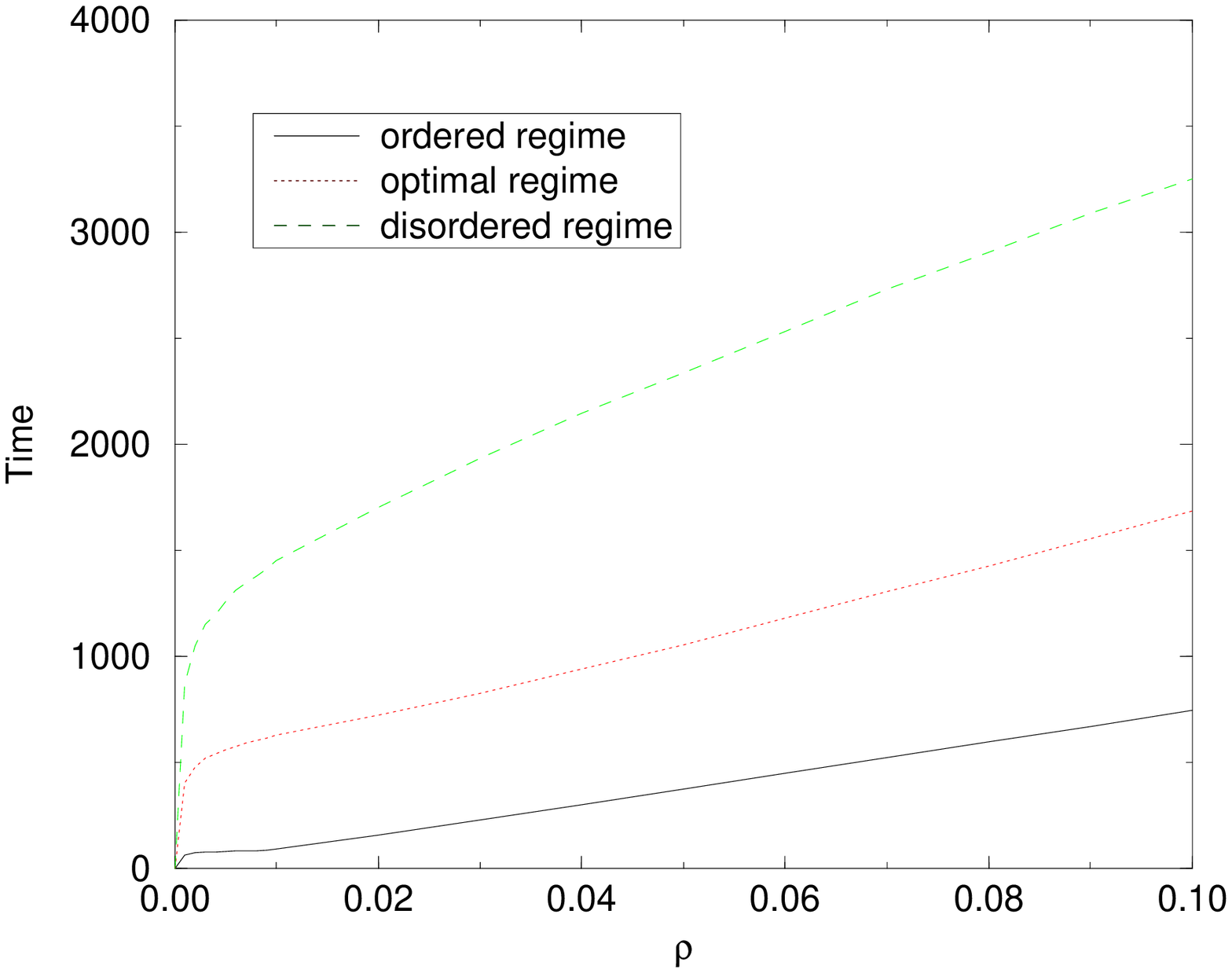}
\end{center}
\caption{Average evacuation times for a large room: 
{\bf (a)} density regime $\rho\in [0,0.9]$; 
{\bf (b)} more detailed look at the low density regime $\rho\in [0,0.1]$.}
\label{densgap2}
\end{figure}

Fig.~\ref{densgap2}(a) shows the influence of an increasing 
particle density for the evacuation times in the three main regimes.
For very small densities one finds a very strong increase of the 
evacuation times with increasing density (fig.~\ref{densgap2}(b)).
If there are only very few particles in the system, which are
distributed randomly over the lattice at $t=0$, they reach the door
nearly independently from each other during the evacuation. Than the
evacuation times are nearly proportional to the largest initial distance 
to the exit.
In this regime diffusion of the dynamic field $D$ provides
advantageous information to the particles (see fig.~\ref{diff} (b)).
In the density regime $\rho\approx 0.005$ a queue in front of the door
begins to form that strongly controls the evacuation time which then
increases nearly linearly with growing density due to clogging.


\subsection{Alternative definitions of the dynamic floor field}
\label{variation}

\begin{figure}[h]
\begin{center}
\includegraphics[width=0.49\textwidth]{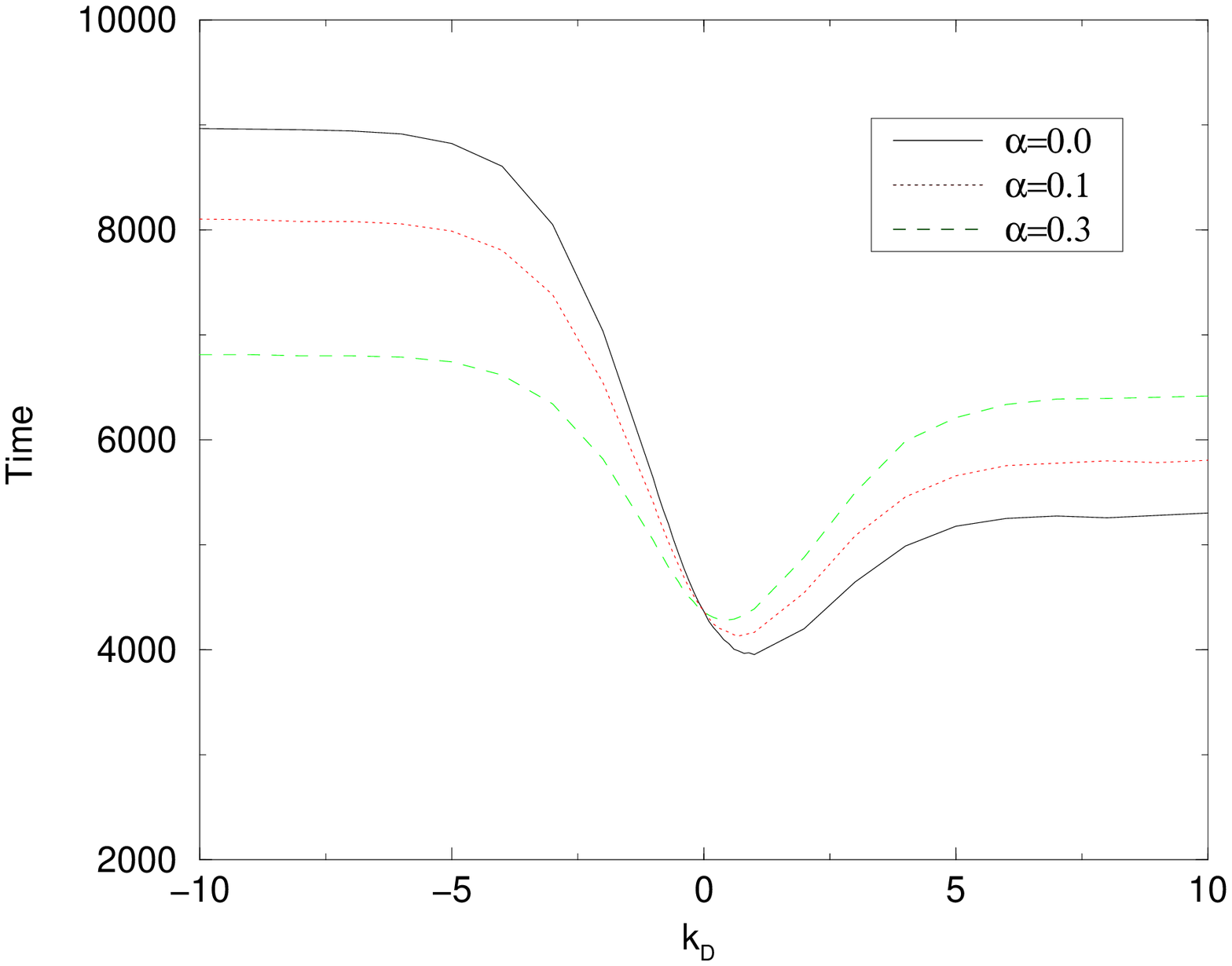}
\includegraphics[width=0.49\textwidth]{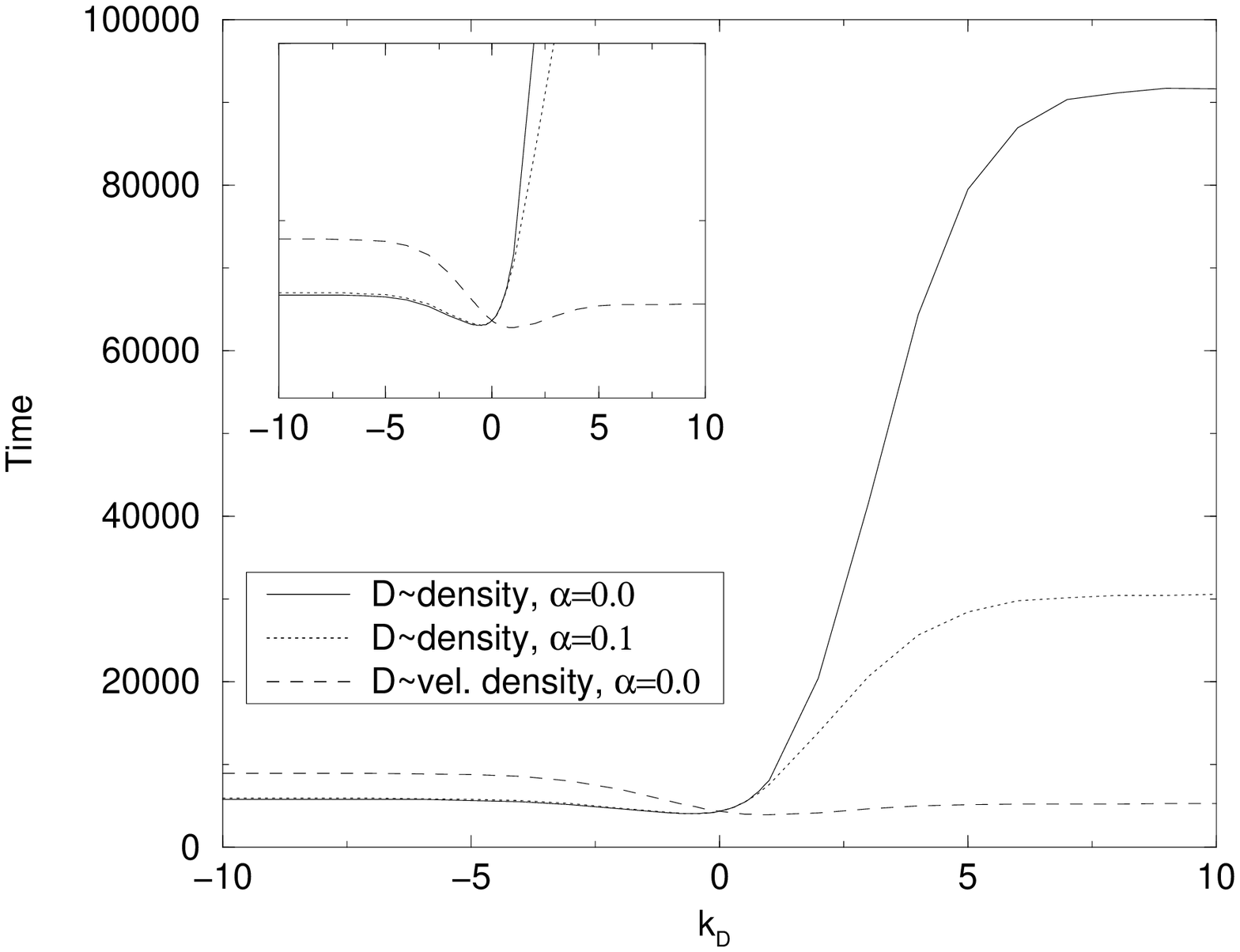}
\end{center}
\caption{Averaged evacuation times for variations of $D$ and $k_D$ for a 
density of $\rho=0.3$ and a lattice of $63\times 63$ sites: {\bf (a)} 
$k_D\in[-10,10]$ for $D$ proportional to the velocity density;
  {\bf (b)} $k_D\in[-10,10]$ for a field $\tilde{D}$ proportional to the 
particle density and $D$ proportional to the velocity density.}
\label{variation_plots}
\end{figure}

In the following we want to give a brief discussion of the consequences of 
several variations and extensions of the definition of the dynamic field 
$D$ and the corresponding coupling parameter $k_D$. 
First we concentrate on the effects of a negative sensitivity parameter 
$k_D < 0$, corresponding to repulsive interactions between the individuals,
on the evacuation times. 
Fig.~\ref{variation_plots}(a) shows the averaged evacuation times $T$ for
a high density of $\rho=0.3$ and weak coupling $k_S=0.4$ to the static 
field for several diffusion parameter values $\alpha$ ($\delta =0.3$ fixed). 
For all diffusion parameters $\alpha$ the averaged evacuation time is
increasing for decreasing sensitivity parameter $k_D < 0$. This is
obvious since the dynamic field $D$ marks the regions of highest
flow in the system (e.g.\ nearby the exit), because of its relation to
the velocity density, and therefore an avoidance of this regions
should lead to an increased evacuation time.  An increasing $\alpha$
value weakens the sharpness of $D$ in the high flow regions and so the
effect of the avoidance of this regions is moderated, i.e.  the
evacuation times decrease for increasing $\alpha$ and $k_D < 0$.
 
\begin{figure}[h]
\begin{center}
\includegraphics[width=0.3\textwidth]{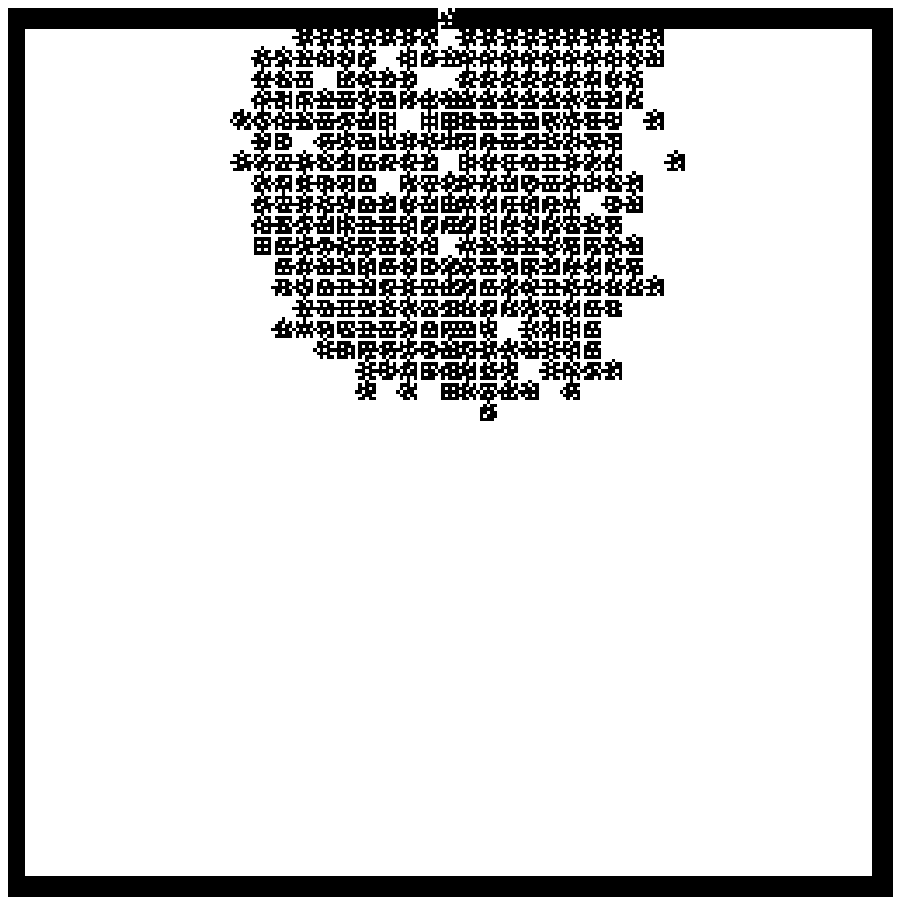}
\includegraphics[width=0.3\textwidth]{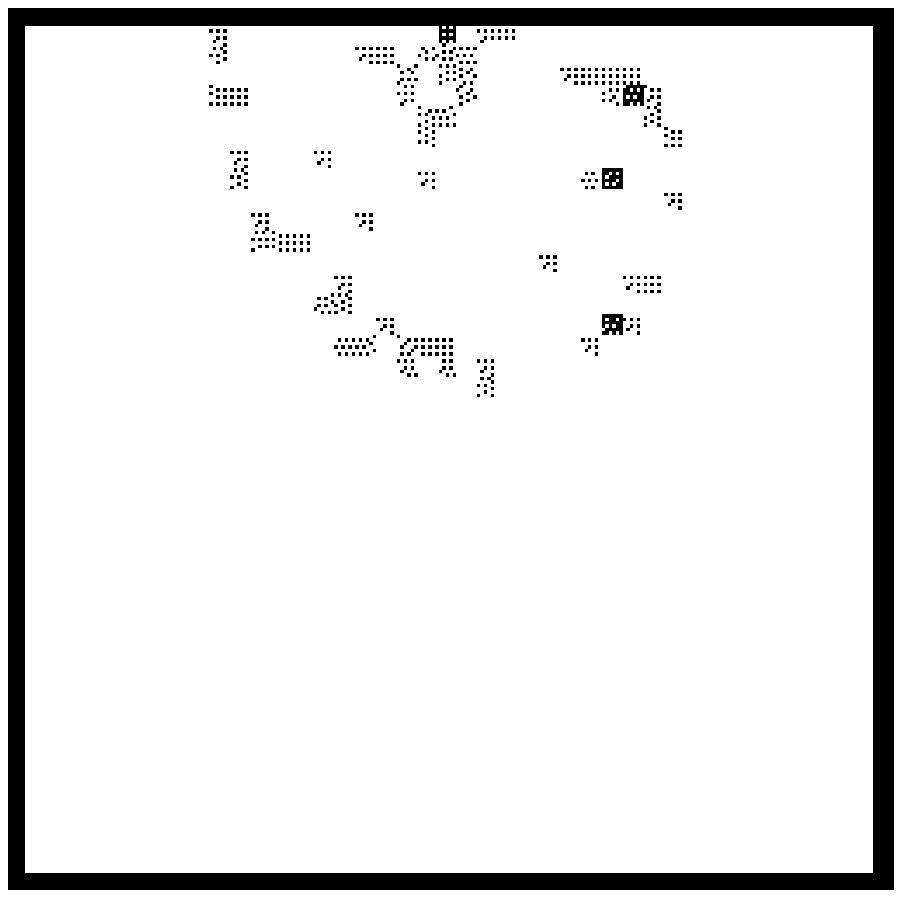}
\includegraphics[width=0.3\textwidth]{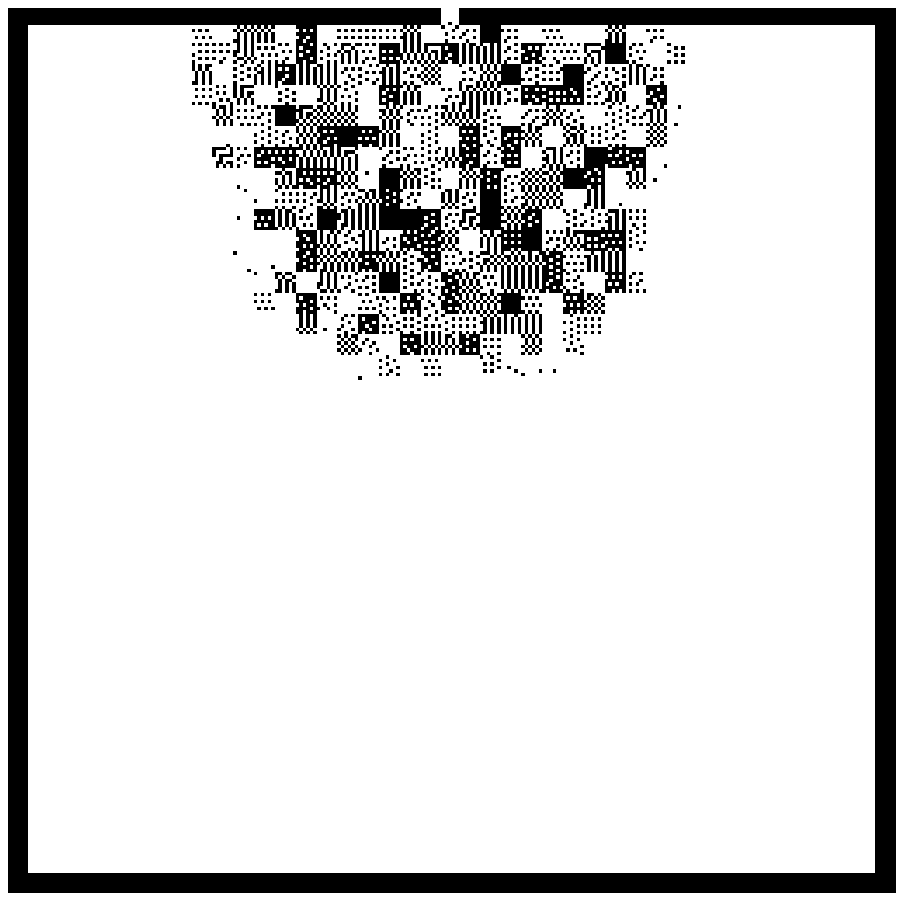}
\end{center}
\caption[]{Snapshot of a simulation and the corresponding dynamic floor 
fields $D$ (for $\alpha=0$, $\delta=0.3$): 
{\bf (a)} typical middle stage of the dynamics of the original model and
{\bf (b)} the corresponding dynamical field $D$ related to the velocity 
density; 
{\bf (c)} dynamical field $\tilde{D}$ for a modified model with 
$\tilde{D}$ related to the particle density.}
\label{denspics}
\end{figure}

More subtle consequences arise for the dynamics of the model if the
definition of the dynamic field $D$ is changed. Up to now we have
considered a dynamic field which is related to the velocity density in
the sense that it is only altered by moving particles (see sec.~\ref{dyn}). 
As an alternative we investigate a field $\tilde{D}$ which is
altered by all particles: in each time step each particle increases the
dynamic field value of its site by one, i.e.\ $\tilde{D}$ is proportional 
to the particle density. Fig.~\ref{variation_plots}(b) shows the
evacuation times for fixed coupling strength $k_S=0.4$ to the static
field, variable $k_{\tilde{D}}\in [-10,10]$ and the two diffusion parameters
$\alpha=0.0$ and $\alpha=0.1$ for this density field $\tilde{D}$ ($\delta
=0.3$ fixed). As a comparison fig.~\ref{variation_plots}(b) shows
again the corresponding times for the velocity density field
$D$ (see fig.~\ref{variation_plots}(a)) with $\alpha=0.0$.  For
positive coupling parameter $k_D>0$, i.e.\ attractive interaction, one
finds monotonously increasing evacuation times for all diffusion parameters
$\alpha$. The time values for different $\alpha$ begin to branch for
$k_D>1$ such that for $k_D>5$ the evacuation time for a
diffusion parameter $\alpha =0.0$ is three times higher than for
$\alpha =0.1$. The reason for that becomes obvious from 
fig.~\ref{denspics}(c) where a two-dimensional visualisation of the
density-dependent dynamic field $\tilde{D}$ is shown.
The darkest shaded sites correspond to regions of highest $\tilde{D}$ values.
Since $\tilde{D}$ is related to the particle density, regions of jamming due
to high particle concentration are signified by high $\tilde{D}$ values 
(e.g.\ a half-circle jamming in front of a door, see 
fig.~\ref{denspics}(a)).
This effect is strongest for $\alpha=0.0$, because a non-vanishing
diffusion parameter only smoothes the field gradients. Therefore an
attractive interaction to such a particle density-dependent field
leads to very high evacuation times due to clustering and herding
effects known from panic situations, especially for vanishing
diffusion of $\tilde{D}$.
 
For negative sensitivity parameter $k_{\tilde{D}}<0$, i.e.\ repulsive 
interaction, one finds a different behaviour. In the region of 
$k_{\tilde{D}}\in\ ]-1,0[$ the evacuation time for the density-dependent 
field case is minimized. For smaller $k_{\tilde{D}}$ values the evacuation 
times increase again monotonically. 
This is similar to the effect of minimized evacuation times for the
velocity density related field $D$ and $k_D\in]0,1[$ (see 
fig.~\ref{fixed_kdks}(b) and inset of fig.~\ref{variation_plots}(b)).
A slight repulsion from regions of high density combined with a weak
guidance through the static ground field $S$ ($k_S=0.4$) should lead
to a larger flow and to smaller evacuation times, as well as a slight
attraction to regions of high flow (inset of
fig.~\ref{variation_plots}(b)). A strong repulsive behaviour supersedes
the directed walk through $S$ and therefore is counterproductive. It
leads again to increasing evacuation times. For $k_D<0$ the evacuation
times for the particle density dependent field $\tilde{D}$ are always 
smaller than for the velocity density dependent field $D$, since a 
repulsion from regions of higher particle density should always be 
more favourable than repulsion from regions of higher flow.
However, fig.~\ref{variation_plots} indicates that a density dependent
field with repulsive interactions behaves in some aspects as a
velocity density dependent field with attractive interactions.

\subsection{Room with two doors}
\label{twodoors}
\begin{figure}[h]
\begin{center}
\includegraphics[width=0.3\textwidth]{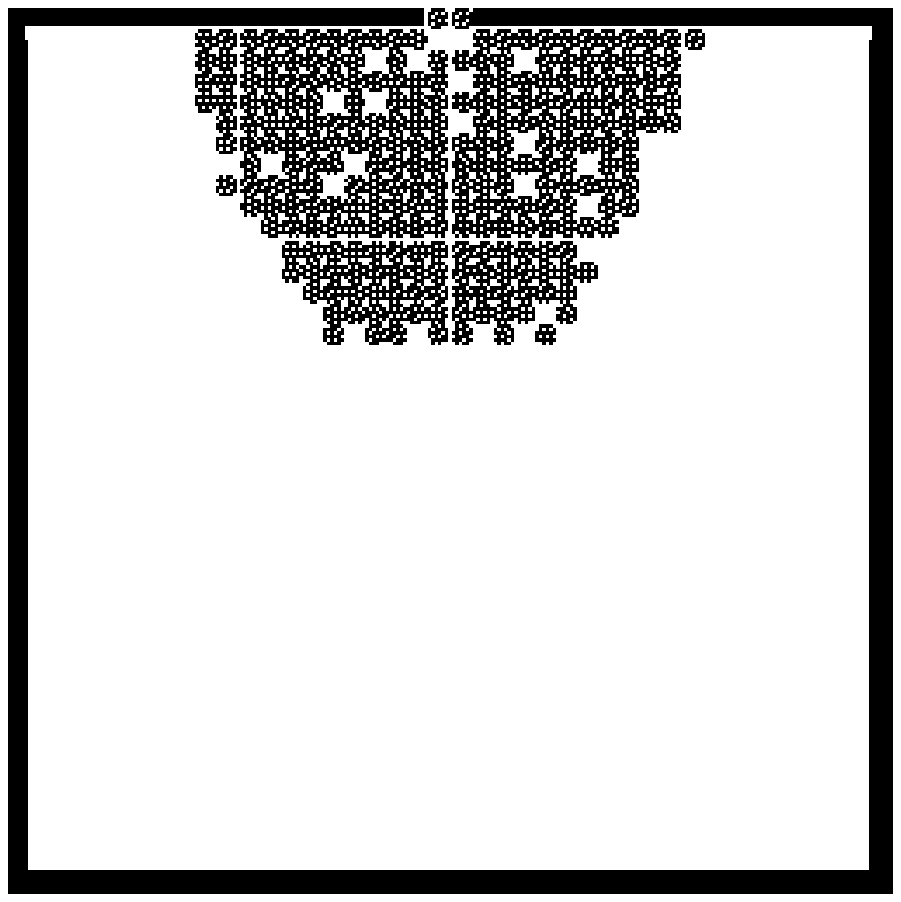}
\includegraphics[width=0.3\textwidth]{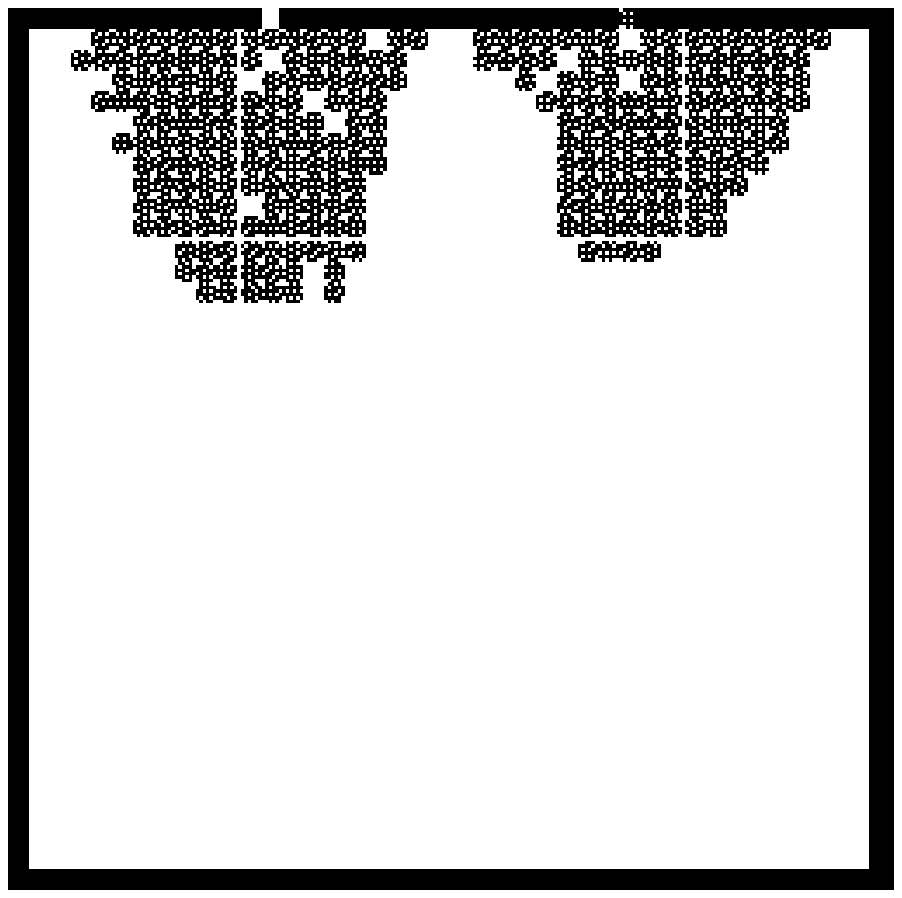}
\includegraphics[width=0.3\textwidth]{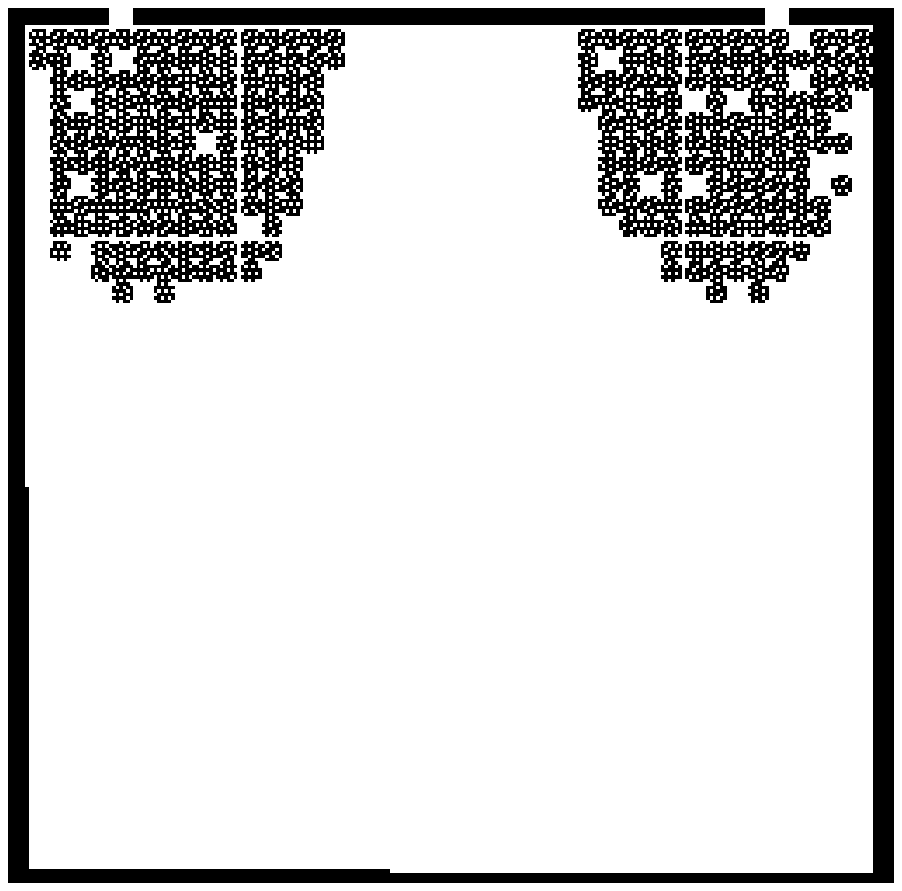}
\end{center}
\caption[]{Typical clogging configurations for a room with two doors and
{\bf (a)} no gap between doors, {\bf (b)} medium gap size, and {\bf (c)} 
doors at the boundaries.}
\label{gap}
\end{figure}

\begin{figure}[ht]
\begin{center}
\includegraphics[width=0.49\textwidth]{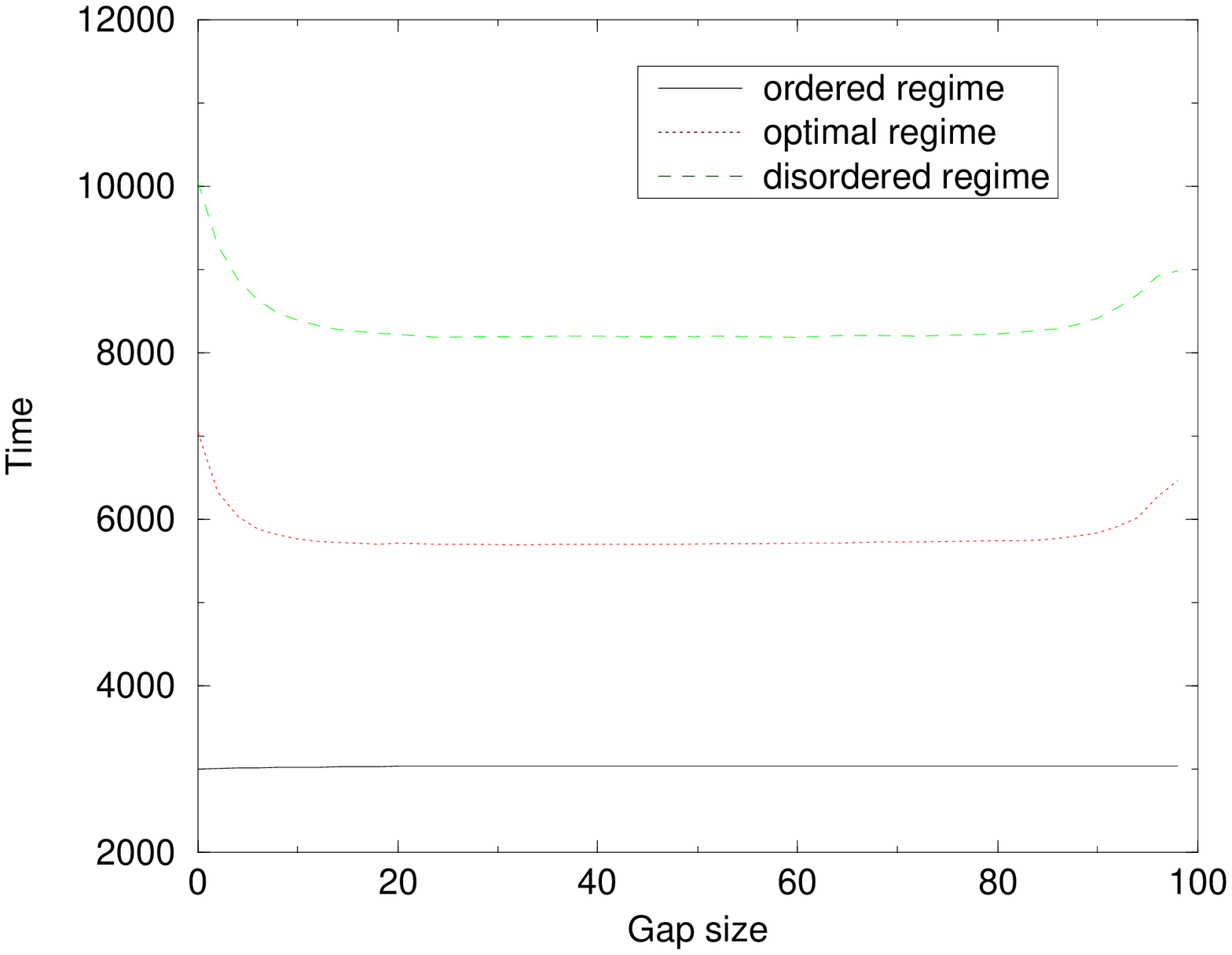}
\includegraphics[width=0.49\textwidth]{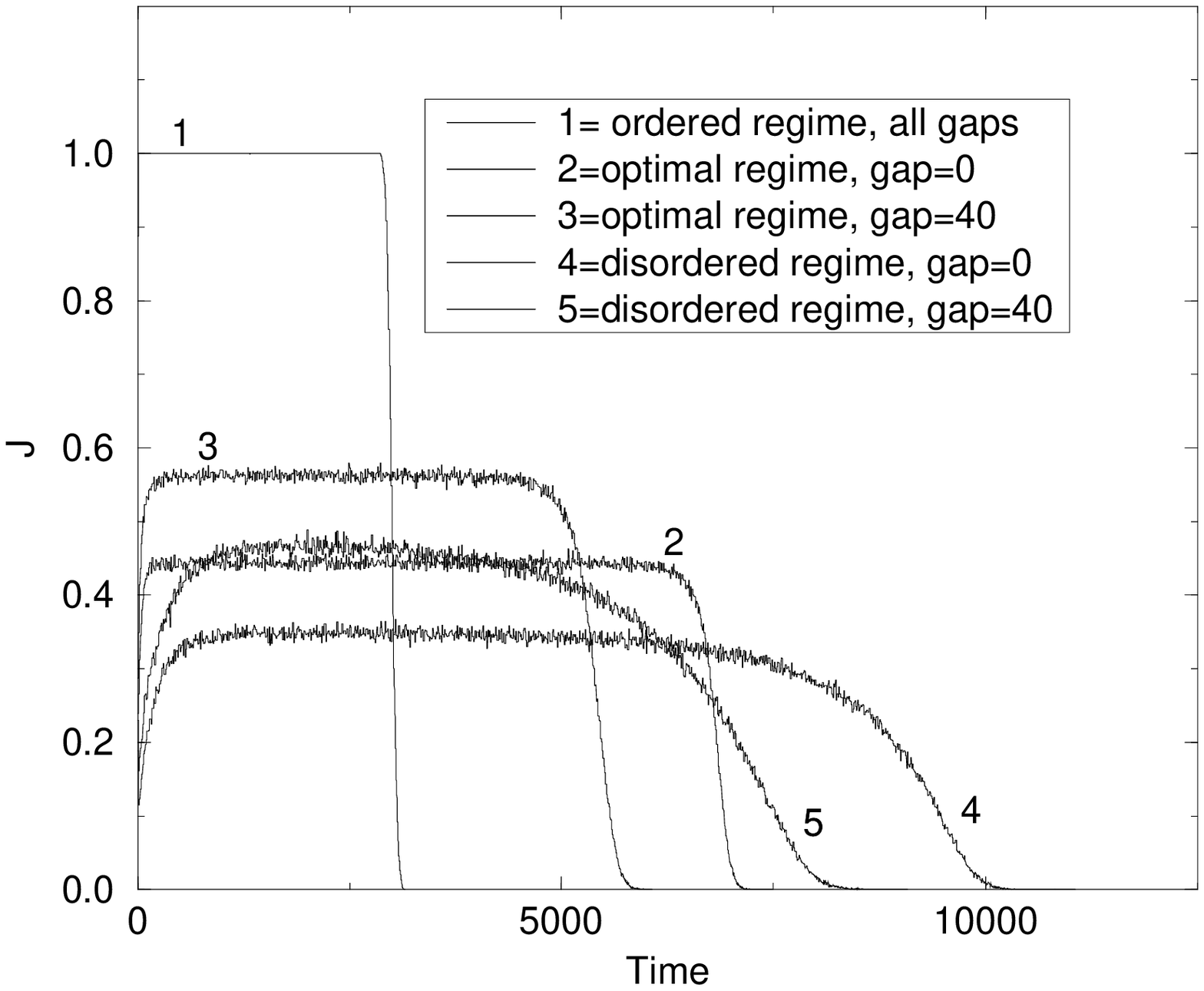}
\end{center}
\caption[]{
{\bf (a)} Evacuation times for all three regimes as function of the gap 
between doors. {\bf (b)} Averaged flows for selected gap sizes for the 
three regimes.}
\label{gap2}
\end{figure}

As a simple example for safety estimations in architectural planing
we investigate how evacuation times change if a gap between 
two doors is increased from zero to a maximal value. We consider 
again an ordinary room with no internal structure of grid size 
$102\times102$. 
We start with one door of width two cells in the middle of one wall.
Then the door is split into two doors of one cell each 
separated by gaps ranging from $2$ to $98$ (fig.~\ref{gap}(b) and (c)). 
Fig.~\ref{gap} shows typical stages of the dynamics for different gap
sizes.
Averaged evacuation times are measured for all three main
regimes introduced in sec.~\ref{sub_sens}. Depending on the regime,
fig.~\ref{gap2}(a) shows a strong influence of the size of 
the gap for small and for large gaps.
In the ordered regime, the evacuation time is almost uninfluenced
by the gap size. The reason is the almost
deterministic motion of the particles which do not interact with 
each other through the dynamic floor field $D$.
In the other two regimes, however, the value of the gap can have
a large influence due to the strong interaction effects through $D$. 
For both the optimal and the disordered regime one finds minimized 
evacuation times for a wide area of the gap length ranging from about 
$20$ to $80$ cells. For small gaps the evacuation times increase due to 
the interactions between the pedestrians. For large gaps the presence
of the side walls has a negative effect (fig.~\ref{gap}(c))
and so again the evacuation times increase.
For intermediate gaps the crowd of pedestrians will be subdivided 
into smaller groups (fig.~\ref{gap}(b)), leading to more favorable 
interactions between them.
This picture is confirmed by the averaged flow values $J$ 
in fig.~\ref{gap2}(b).
A gap size of $40$ cells between the doors leads to maximised flow values 
in the optimal and disordered regime.

\section{Conclusions}

We have studied simple evacuation processes using a recently introduced
stochastic cellular automaton for pedestrian dynamics which implements
interactions between individuals using an idea similar to chemotaxis. 
Due to its simplicity the model allows very high simulation speeds and 
is very well suited for the optimization of evacuation procedures even
in complex situations.

We have focussed on studying evacuation times
in a very simple evacuation scenario. The main purpose
was to elucidate the influence of the various model parameters,
especially of the coupling strengths $k_S$ and $k_D$ to the static and 
dynamic floor fields, to facilitate their interpretation.

Most important are the coupling parameters $k_D$ and $k_S$ to the
dynamic and static floor fields, respectively. $k_S$ is a measure
for the knowledge of the individuals about the geometry, especially
the location of the exits. For dominating coupling to the static
field the pedestrians will choose the nearest exit without much
detour. For $k_S\to 0$, on the other hand, they have no knowledge
at all and will find the doors just by chance. 

The dynamic floor field $D$ is a measure for the velocity density
of the pedestrians.
The parameter $k_D$ controls the herding behaviour. For dominating 
coupling to the dynamic field the pedestrians have a strong tendency
to follow in the footsteps of others, e.g.\ because they hope
that others have more knowledge about the location of exits.
Such a behaviour is relevant for panic situations where this
herding tendency becomes important and has been observed empirically
\cite{HePED,panic}. In fact, the ratio $k_D/k_S$ has very similar
effects as the panic parameter introduced in \cite{panic}.

The investigation of alternative definitions of the dynamic floor
field has shown that the most realistic results are obtained for 
the velocity density dependent field with repulsive interactions
as used in the original definition of the model. However, 
a field related to the particle density with repulsive interactions
behaves in some aspects very similarly.

An important result of our investigations is that for achieving 
optimal evacuation times a proper combination of herding behaviour
and use of knowledge about the surrounding is necessary.
Then through cooperative behaviour of the individuals evacuation
times can be minimized.

We want to emphasize that here only a very simple scenario has been studied
in order to identify the different regimes of the model. For realistic
applications a procedure to determine the coupling parameters is needed.
This work is in progress and results will be published elsewhere
\cite{KNS,aki}. 
In the Appendix one possible idea is discussed (see also \cite{ourpaper}). 
Furthermore the effects of disorder (e.g.\ different individuals $j$
having different couplings $k_S^{(j)}$ and $k_D^{(j)}$) are important.
Another interesting open question concerns the details of the dynamics
of evacuation processes in panic situations. We have assumed, as is
also done for the panic parameter in \cite{panic}, that the coupling 
constants do not change during the process. However, in many realistic
situations it might be more appropriate to use a panic parameter
which increases with time. We leave this important problem for
future study.

It is suprising that the properties of the model investigated here
are in many respects very similar to the social-force model \cite{social}
although the interactions are very different. In our approach pedestrians
interact with the velocity-density through an attractive coupling
whereas the social-force model uses a repulsive density-density interaction.
It would be interesting to get a deeper understanding of these
similarities.



\section*{Acknowledgment}
\noindent

We like to thank K.\ Nishinari, C.\ Burstedde, F.\ Zielen and D.\ Helbing
for useful discussions.


\appendix
\section{Construction of the static floor field $S$}
\label{CoS}

The values $S_{ij}$ for the static floor field for sites $(i,j)$ for
the geometries used in this paper can be calculated in the following
way: The lattice shall be surrounded by obstacle cells, which are not
traversable, except for a number of door cells at
$\{(i_{T_1},j_{T_1}),\cdots,(i_{T_k},j_{T_k})\}$. The pedestrians can
only leave the room through these door cells.  The explicit values of
$S$ in the examples studied here are than calculated with a 
distance metric:
\begin{equation}
\label{ped_feldformel}
    S_{ij}= \min_{(i_{T_s},j_{T_s})}\left\{\max_{({i}_l,{j}_l)}
      \left\{\sqrt{(i_{T_s}-{i}_l)^2+(j_{T_s}-{j}_l)^2}\right\}-
      \sqrt{(i_{T_s}-i)^2+(j_{T_s}-j)^2}\right\}\,.
\end{equation}
This means that the strength of the static floor field depends on the
shortest distance to an exit. $\max_{({i}_l,{j}_l)} \left\{
\sqrt{(i_{T_s}-{i}_l)^2+(j_{T_s}-{j}_l)^2} \right\}$, 
where $({i}_l,{j}_l)$ runs over all cells of the lattice,
is the largest distance of any cell to the exit at $(i_{T_s},j_{T_s})$.
This is just a normalisation so that the field values increase with
decreasing distance $\sqrt{(i_{T_s}-i)^2+(j_{T_s}-j)^2}$ to an exit and 
is zero for the cell farthest away from the door.

This is only one possible representation of the explicit calculation of the
static floor field $S$. Alternative constructions of $S$ do not change
the qualitative results of the investigations, provided that the
strength of the static floor field is increased in the direction to
the exit \cite{aki}. The use of a Manhattan metric is one possible
example for a successful construction of $S$ for more complex
geometries \cite{KNS,aki}.

Once the floor field is specified from the geometry of the problem
the coupling constant $k_S$ has to be determined such that a quantitative
agreement with experimental data is obtained. Currently we are 
investigating different strategies \cite{KNS,aki}. A simple method
would be to determine $k_S$ such that the observed velocity
and velocity fluctuations of the motion of a single pedestrian are 
reproduced (see \cite{ourpaper}). 



\end{document}